{%revtex4}
\documentclass[showpacs,amssymb,floatfix,pre,aps,notitlepage,preprint]{revtex4-1}

\usepackage{dcolumn}   % needed for some tables
\usepackage{bm}        % for math
\usepackage{amssymb}   % for math

\usepackage{amssymb,amsfonts,bm}
\usepackage{graphics,graphicx}
\usepackage{colordvi,amsmath,epsfig,color}
\usepackage[activeacute,english]{babel}
\bibliographystyle{prsty}
\providecommand{\abs}[1]{\lvert#1\rvert}

\usepackage{appendix}
\usepackage{dcolumn}
\usepackage[mathscr]{euscript}
\usepackage{amsmath}
\usepackage{xcolor}
\usepackage{physics}

\usepackage{subcaption}
\usepackage{hyperref}
\usepackage{soul}

\newcommand\gammaf[1]{ \; \Gamma\left(#1\right)}

\newcommand{\J}{\mathcal{J}[f|f]}
\newcommand{\Jz}{\mathcal{J}_z[f|f]}
\newcommand{\Jt}{\mathcal{\widetilde J}[f|f]}

\begin{document}

\title{Confined granular gases under the influence of vibrating walls}
%\input author_list.tex       % D0 authors (remove the first 3 lines
                             % of this file prior to submission, they
                             % contain a time stamp for the authorlist)
                             % (includes institutions and visitors)

\author{M. Mayo}
\author{J. C. Petit}
\affiliation{F\'{\i}sica Te\'{o}rica, Universidad de Sevilla,
Apartado de Correos 1065, E-41080, Sevilla, Spain}
\author{M. I. Garc\'ia de Soria}
\author{P. Maynar}

\affiliation{F\'{\i}sica Te\'{o}rica, Universidad de Sevilla,
Apartado de Correos 1065, E-41080, Sevilla, Spain}
\affiliation{Institute for Theoretical and Computational
  Physics. Facultad de Ciencias. Universidad de Granada, E-18071,
  Granada, Spain.%\\This line break forced% with \\
}%

\date{\today}

\begin{abstract}
The dynamics of a system composed of inelastic hard spheres or disks
that are confined between two parallel vertically vibrating walls is
studied (the vertical direction is defined as the direction
perpendicular to the walls). The distance between the two walls is supposed to be larger
than twice the diameter of the particles so that the particles can
pass over each other, but still much smaller than the dimensions of
the walls. Hence, the system can be considered to be
quasi-two-dimensional (quasi-one-dimensional) in the hard spheres
(disks) case. For dilute systems, a closed evolution equation for the
one-particle distribution function is formulated that takes into
account the effects of the confinement. Assuming the system is
spatially homogeneous, the kinetic equation is solved approximating
the distribution function by a two-temperatures (horizontal and
vertical) gaussian distribution. The obtained evolution equations for the
partial temperatures are solved, finding a very good agreement with 
Molecular Dynamics simulation results for a wide range of the
parameters (inelasticity, height and density) for states whose
projection over a plane parallel to the walls is
homogeneous. In the stationary state, where the energy lost
in collisions is compensated by the energy injected by the walls, the 
pressure tensor in the horizontal direction is analyzed and its
relation with an instability 
of the homogenous state observed in the simulations is discussed. 

\end{abstract}

\maketitle

\title{Confined gas}

\author{Authors here}

%\address{Física Teórica, Universidad de Sevilla, Apartado de Correos 1065,  E-41080, Sevilla, Spain}
%\ead{submissions@iop.org}
\vspace{10pt}
%\begin{indented}
%\item[]December 2022
%\end{indented}

%\noindent{\bf{Keywords}}: XXXXXX, YYYYYYYY, ZZZZZZZZZ
%\maketitle

\section{Introduction}\label{SecI}

Granular gases are systems composed of macroscopic particles which do
not conserve kinetic energy when they collide. As a
consequence, they are intrinsically out of equilibrium and have a very
rich phenomenology \cite{g03, at06, brilliantov, garzo19}, so they can be
considered as a proving ground for 
nonequilibrium statistical mechanics. In the last years, many studies
have been focused on a thin vertical granular system that is vertically vibrating
\cite{peu02, ou05, mvprkeu05, cms12, nrtms14, cms15, gs18}. The reason is
that, for a wide class of initial 
conditions, when the system is observed from above (or below), a
homogeneous stationary state is reached in the long time limit. In
this state, the energy lost in collisions is compensated by the energy
injected by the walls. The fact that the system is out of equilibrium
and, still, spatially homogeneous, is very peculiar. In effect, in
molecular systems the out of equilibrium character is usually linked
to the presence of gradients. For granular systems, this is also the
case as can be seen from the hydrodynamic equations
\cite{bdks98}. For this particular setup, the
gradients appear in the vertical direction and can be neglected if the
height of the system is small enough. Interestingly, when varying
the values of some parameters, as the amplitude of the oscillations or
the total density, the system reaches a different state in which a
solid-like phase is surrounded by a hotter dilute gas. The origin of
this phase transition has been extensively studied (see, for
example, the reviews \cite{mvprkeu05, gs18}) being the ingredient that
triggers the instability a negative compressibility in the horizontal
direction \cite{brs13}. To be precise, the horizontal stationary
pressure tensor is a monotonically decreasing function of the mean
density. Then, if a dense region is spontaneously developed, it has
associated a lower pressure attracting more particles and triggering
the instability. 

Many of the above studies have concentrated in the ultra-confined case
in which the height of the 
system is smaller than twice the diameter of the particles, in such a
way that particles can not jump over each other and the system can be
considered quasi-two-dimensional (Q2D). One of the simplest models
used for the theoretical studies
for this Q2D system is an ensemble of inelastic spheres confined
between two flat planes that inject energy into the system in the
vertical direction \cite{mgb19, mgb19b, mgb22}. Gravity is not
consider and by vertical direction we mean the direction perpendicular
to the walls. The energy injection
is the corresponding to an 
elastic wall that vibrates sinusoidally with amplitude $A$ and
frequency $\omega$ in the double limit $A\to 0$, $\omega\to\infty$
with $A\omega=v_p$. This model is particularly appealing as the wall
can be considered to be always at the same place simplifying the
theoretical analysis. If a particle collides with the bottom (top)
wall, it sees the wall moving upwards (downwards) with velocity
$v_p$. In some studies the bottom wall vibrates, the top wall
being at rest \cite{mgb19, mgb19b}, while in others both walls
vibrate with the same velocity \cite{mgb22}.  
A kinetic equation has been proposed that describes very well the
dynamics of the system in the dilute limit\cite{mgb19}. The equation
is a Boltzmann-like equation in which 
the collisional contribution takes into account that the only possible
collisions are the ones with an orientation such that the two
particles are inside the container. Let us stress that 
the equation was previously introduced for elastic hard spheres (in
this case the walls are at rest and the collisions of particles with
the wall do not inject energy), finding also an excellent agreement
between the theoretical results and Molecular Dynamics (MD)
simulations \cite{bmg16, bgm17, mbgm22}. The formalism can be generalized to
arbitrary confinement and for moderate densities at the Enskog level
\cite{mgb18}. Coming back again to the inelastic case, the 
obtained equation of state has a negative compressibility, compatible
with the presence of the instability. Moreover, if a hydrodynamic
description in the horizontal direction is assumed, it is found that, 
depending on the horizontal size of the system, the homogeneous state is
stable or unstable \cite{mgb19b}. The ingredient that makes the system unstable is,
as pointed above, the negative compressibility, while the one that
stabilizes it is heat conduction. Depending on which of the two
ingredients dominate, the system is stable or unstable. Let us remark
that an excellent agreement between the theoretical prediction and MD simulations for
the critical size is found \cite{mgb19b}.   

Other studies consider Q2D granular systems but without the
limitation on the height. The system is confined in the vertical
direction being the horizontal size much larger
than the vertical one, but it is not ultraconfined in the sense that
the height can be larger than twice the 
diameter of the particles (so that the grains can pass over each
other) \cite{rcbhs11, crbhs12, cwbhsm16}. In these works, the simulations and
experiments are performed for dilute systems (the packing fraction is
of the order of 0.05) and a similar phenomenology is found: for a wide
range of the parameters a homogeneous stationary state is reached 
while, when varying the vibrating amplitude or the density, an
aggregate can be developed. Moreover, it seems that the instability is
also triggered by a negative compressibility. Let us point out that
the simulations were performed without gravity, neglecting the
particle-wall friction and rotational effects, 
obtaining a very similar phase diagram than in the experiments. 
In this paper, the objective is to start the theoretical study of this
kind of systems assuming a kinetic theory description. Taking into
account the simplified conditions of the simulations performed in
\cite{rcbhs11, crbhs12, cwbhsm16}, the model used in \cite{mgb19,
  mgb19b} will be considered, i.e., inelastic hard spheres or disks
confined between two sawtooth walls, but without the
restriction on the height. We will consider that the two walls
vibrate in the sawtooth way described above because, when the system
is vertically agitated, both walls inject energy into the system. 
The main questions to address are to investigate the  accuracy of a
kinetic theory 
description and the possibility of obtaining an equation of state in
the horizontal direction. If this is the case, the concrete dependence
of the horizontal pressure on the density is essential to understand
microscopically the instability. 

The paper is organized as follows: in Sec. \ref{SecII} the model is
introduced and the corresponding kinetic equation is derived in the
low density limit. Everything is done for arbitrary spatial dimension,
$d$, i.e. $d=3$ for hard spheres confined between two flat planes and $d=2$ for
hard disks confined between two lines. The dynamics of the spatially
homogeneous case is studied in 
Sec. \ref{SecIII}, where the kinetic equation is approximately solved
assuming that the distribution function is a gaussian with two
different temperatures, the horizontal and the vertical
temperatures. The stationary solutions are studied and its stability
is analyzed. In Sec. \ref{SecIV} the theoretical results are compared with
MD simulation results finding a very good agreement for a wide range
of the parameters. Finally, Sec. \ref{SecV} contains a short summary
of the results and some of the possible implications are
discussed. Details of the calculations are presented in the
Appendices.  

\section{The model}
\label{SecII}

We consider an ensemble of $N$ smooth inelastic hard spheres
($d=3$) or disks ($d=2$) of mass 
$m$ and diameter $\sigma$ that are confined between  two
parallel plates of area $A$ (two square walls of area $A\equiv
L^2 $ in $d =3$ or two lines of length $L$ in  $d =2$). The walls are located
at $z = 0$ and $z = H$ respectively and the unitary normal
vector to the walls is ${\widehat{ \bm e}}_{z }$. It is assumed that
$H\ge 2\sigma$ (the case $H\le 2\sigma$ was already studied in
\cite{mgb19}). 
Since gravity is neglected and no other external force is applied,
particles move freely between consecutive collisions. When there is a
binary encounter between two particles of velocities $\bm{v}$ and
$\bm{v}_1$, the velocities of the particles after the collision, $\bm{v}'$ and
$\bm{v}_1'$, are given by 
\begin{eqnarray}
\label{2.1}
{\bm v}^{\prime} &\equiv&  b_{{\widehat{\bm\sigma}}} {\bm v} = {\bm
  v}+ \frac{1+\alpha}{2}\left( {\bm g} \cdot \widehat{\bm
  \sigma} \right) \widehat{\bm \sigma}, \\
\label{2.2}
{\bm v}^{\prime}_{1} &\equiv& b_{\widehat{\bm\sigma}} {\bm v}_{1} = {\bm v}_{1}- \frac{1+\alpha}{2}\left( {\bm g} \cdot \widehat{\bm \sigma} \right) \widehat{\bm \sigma},
\end{eqnarray}
where 
${\bm g} \equiv {\bm v}_{1} - {\bm v}$ is the relative
velocity between particles before the collision 
and $\widehat{\bm\sigma}$ is an unitary vector directed along the line
joining the centers of the particles at contact away from particle
with pre-collisional velocity $\bm{v}$. We have also introduced the
operator $b_{\widehat{\bm\sigma}}$ that changes pre-collisional 
to post-collisional velocities. The coefficient of restitution,
$\alpha$, varies in the range of $0 
\leq\alpha \leq1$ and  
is assumed to be constant, being $\alpha=1$ the elastic case. We
always consider  inelastic systems, $\alpha < 1 $, and periodic
boundary conditions in the horizontal directions. The bottom wall
located at $z=0$ is modeled by a sawtooth wall with velocity $v_p$.
Each time a particle collides with it, the particle  always sees  the
wall moving along the vertical direction with positive velocity ${\bm
  v}_p= v_p\widehat{\bm  e}_z$ . Thus, for a particle with pre-collisional velocity $\bm{v}$,
the post-collisional velocity is given by  
\begin{equation}
b_{\downarrow}{\bm v}\equiv {\bm v} + 2(v_p - v_{z})\widehat{\bm
  e}_{z}, 
\label{ecu3}
\end{equation}
where $b_{\downarrow}$ is an operator that changes pre-collisional to
post-collisional velocities (the symbol $\downarrow$ in the operator means that the
particle collides with the bottom wall). 
Note that the collision between a particle and the bottom wall only
takes place if $v_z<0$. The top wall is also modeled as a sawtooth
wall with velocity ${\bm v}_p= -v_p\widehat{\bm e}_z$. Under the same
assumptions, we can establish the post-collisional velocities of a
particle with the top wall as  
\begin{equation}
b_{\uparrow}{\bm v} \equiv  {\bm v} - 2(v_p + v_{z})\widehat{\bm e}_{z},
\label{ecu4}
\end{equation}
\begin{figure}[t!]
\centering
\begin{subfigure}{0.6\textwidth}
    \centering \includegraphics[width=8cm]{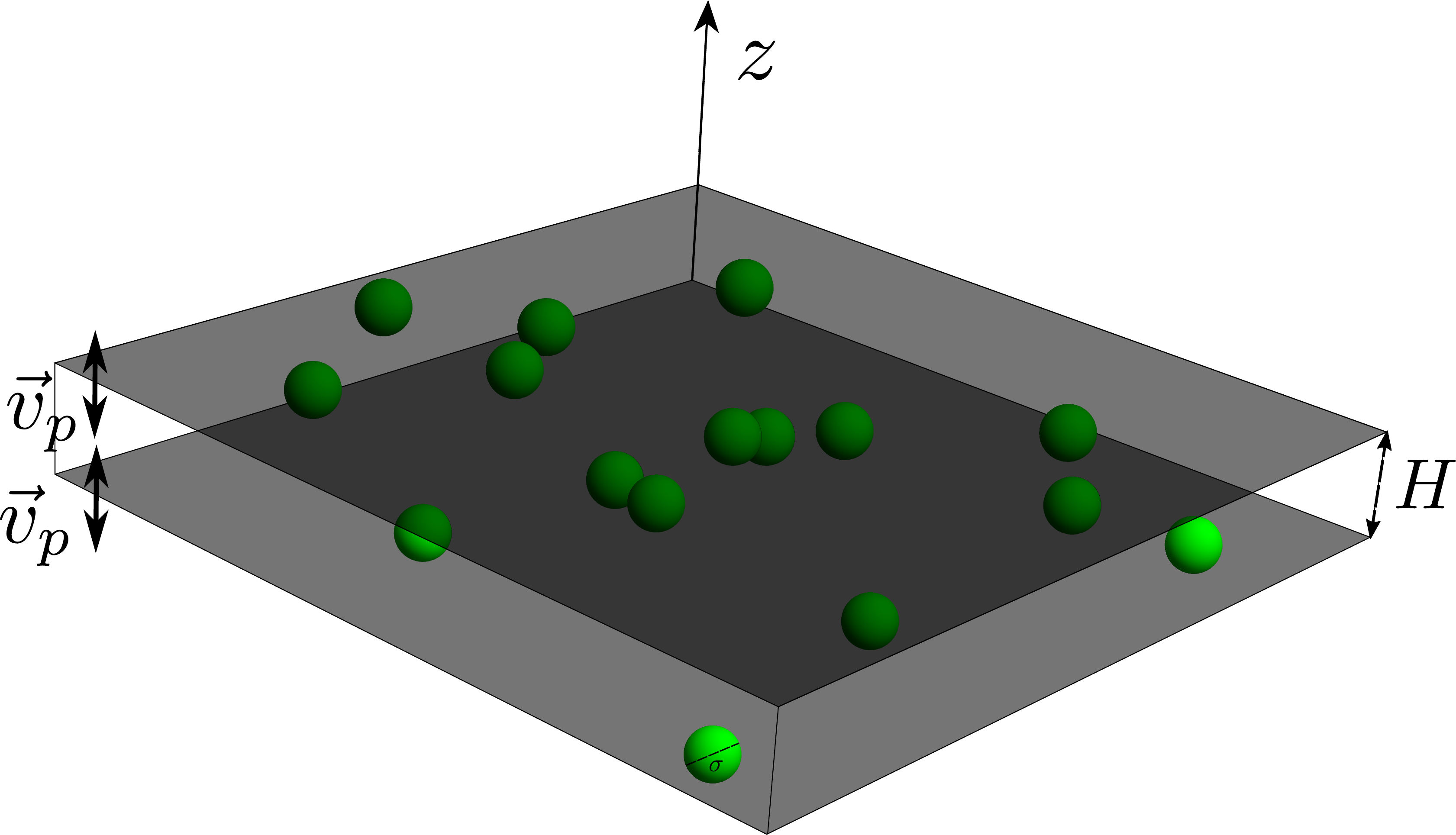}
\caption{}
    \label{fig1:first}
\end{subfigure}
\begin{subfigure}{0.8\textwidth}
  \centering \includegraphics[width=8cm]{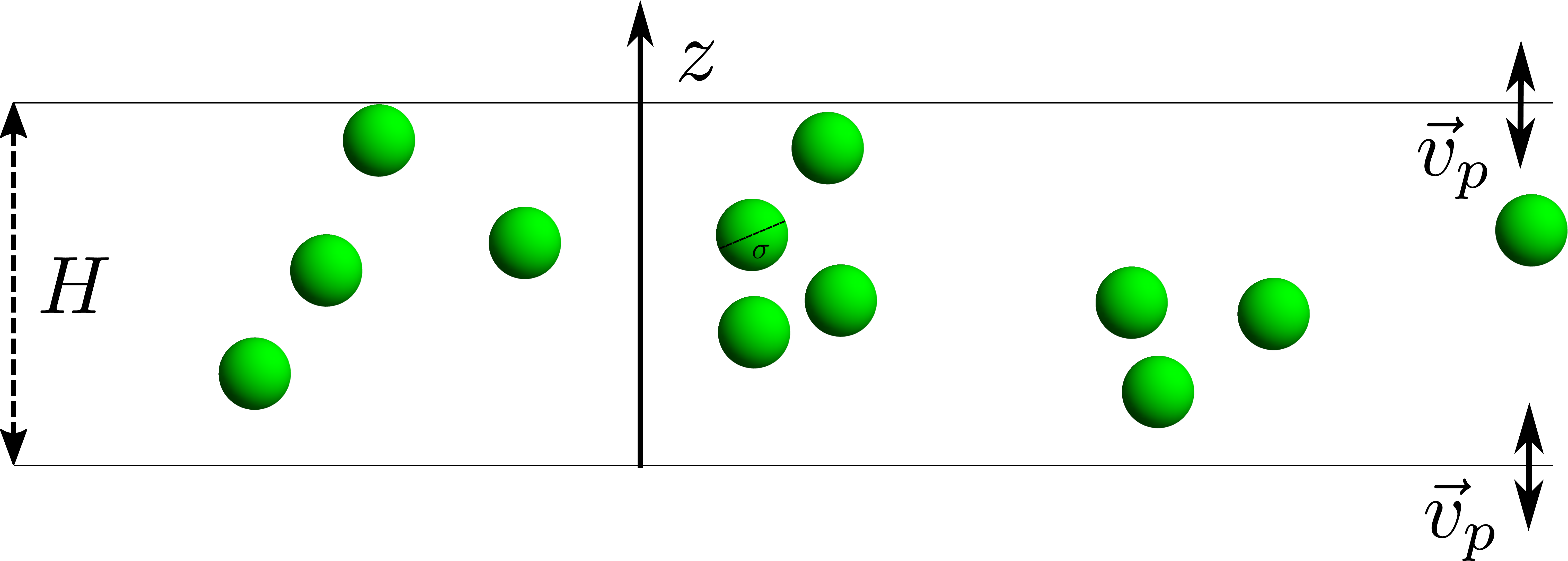}
\caption{}
    \label{fig1:second}
\end{subfigure}        
 \caption{Schematic representation of the model for $d=3$
   (\subref{fig1:first}), and for $d=2$ (\subref{fig1:second}).}
\label{fig1}
\end{figure}
with $v_z>0$. The symbol $\uparrow$ in the operator means that the
particle collides with the top wall. As it was mentioned above, this
simple model for a vibrating wall 
arises for a wall that moves sinusoidally with amplitude $A$, angular
frequency $\omega$, in the limit $A\to 0$, $\omega\to\infty$ with
$A\omega=v_p$. Note that both wall-particle
collisions always inject energy into the system, while the tangential
component of the velocity remains constant. Consequently, the linear
momentum in the horizontal direction is a constant of motion. Let us
remark that in the model the value of $v_p$ can be scaled. If the
values of the walls are changed by a constant factor, i.e, $v_p \to K
v_p$ with $K>0$ and the velocities of the particles are modified by
the same factor, the sequence of collisions is the same due to the
linear character of the collision rule \cite{mgb19}. A schematic
representation of the model for $d = 3$ and $d =2$ is plotted in
Figs. \ref{fig1:first} and \ref{fig1:second} respectively. 

\begin{figure}[t!]
  \centering \includegraphics[scale=0.58]{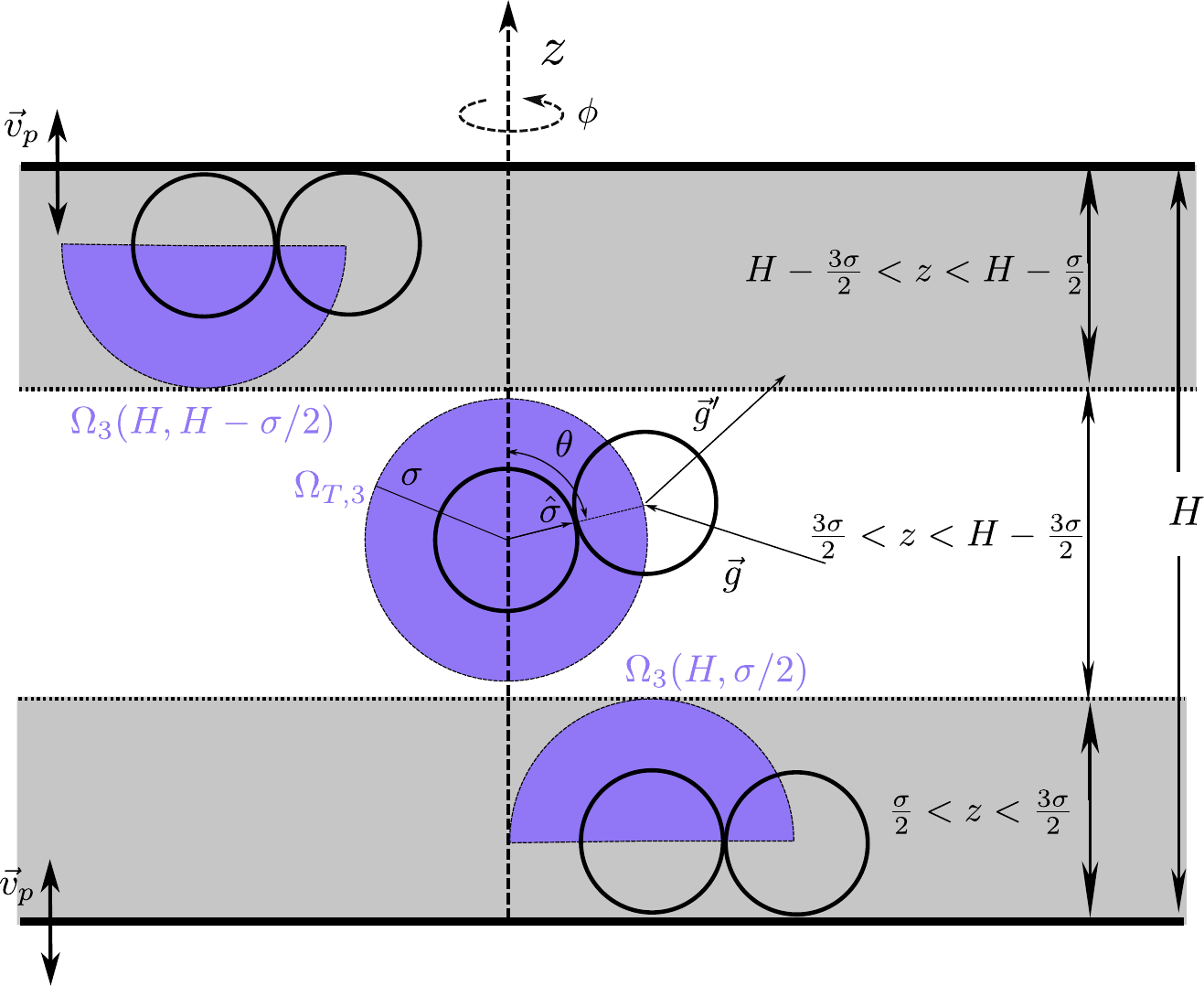}
 \caption{(Color online) Schematic diagram of the possible angles of
collisions in a hard sphere system (purple region). It can be
appreciated that, depending on $z$, three parts in 
the system can be distinguished: one ``bulk'' part where all collisions are
possible and two ``boundary'' parts where the orientation of the
collisions is restricted due to one of the walls. The ``boundary''
part is the region between the dot line and the closest wall (grey
region).    }
\label{fig2}
\end{figure}

In the following, we will assume that the system is in the dilute limit
and we will focus on obtaining a kinetic formulation of the model.  
We assume a closed description of the system in terms of the
one-particle distribution function, $f({\bm r} ,{\bm v}
,t)$. Following standard arguments
\cite{resibois1977classical,dorfman2021contemporary,mclennan1989introduction},
the time evolution equation for the distribution function can be
written as a sum of a free-streaming 
contribution and two parts corresponding to collisions:
particle-particle and particle-wall collisions. In the low density
limit, the particle-particle collisional term can be written as a
functional of the one-particle distribution function by
assuming \textit{Stosszahlansatz} (molecular chaos), i.e.,  
the velocities of colliding particles are uncorrelated and the
evolution equation reads 
\begin{equation}
\left(\frac{\partial}{\partial t} 
+ \bm{v}\cdot \frac{\partial}{\partial \bm{r}} \right) f(\bm{r}, \bm{v}, t) 
= \Jz + L_W f(\bm{r},\bm{v},t).
\label{ecu5}
\end{equation}
Here $\Jz$ is the particle-particle collision term  
\begin{equation}
\Jz =  \sigma^{d-1}\int \dd \bm{v}_{1} \int_{\Omega_d(H, z)} \dd
\widehat{\bm\sigma} \,
|\bm{g}\cdot\widehat{\bm
  \sigma}|\left[\Theta(\bm{g}\cdot\widehat{\bm\sigma})
  \alpha^{-2}b_{\sigma}^{-1} -
  \Theta(-\bm{g}\cdot\widehat{\bm\sigma})\right]f(\bm{r}_1,
\bm{v}_{1}, t) f(\bm{r}, \bm{v}, t),  
\label{ecu6}
\end{equation}
where we have introduced the Heaviside step function, $\Theta$, ${\bm
  r_1} \equiv \bm{r}+ \sigma\widehat{\bm\sigma}$, and the
inverse operator $b_{\widehat{\bm \sigma}}^{-1}$ that changes all
velocities appearing to its right by the pre-collisional
velocities  
\begin{eqnarray}
\bm{v}^{*} &\equiv& b_{\widehat{\bm\sigma}}^{-1}\bm{v} = \bm{v} + \frac{1+\alpha}{ 2\alpha} (\widehat{\bm \sigma}\cdot\bm{g})\widehat{\bm \sigma},\\
\label{ecu7}
\bm{v}_{1}^{*} &\equiv& b_{\widehat{\bm\sigma}}^{-1}\bm{v}_{1} =
\bm{v}_{1} - \frac{1+\alpha}{2\alpha} (\widehat{\bm
  \sigma}\cdot\bm{g})\widehat{\bm\sigma}. 
\label{ecu8}
\end{eqnarray}
The integral over the $d$-dimensional solid angle is taken over the
allowed scattering angles, $\Omega_d(H, z)$, that depends on $H$ and $z$ due to the
walls. For $d=3$, working in spherical coordinates being $\theta$ and $\phi$ the
polar and azimuthal angles respectively, the domain $\Omega_3(H, z)$ can be
parametrized as follows 
\begin{align}\label{ecu9} \Omega_3(H, z) \equiv \left\lbrace 
\begin{aligned}
    \begin{split}
           \{ (\theta, \phi)|\;\theta\in (0, \pi/2+b_1(z)), \phi\in
           (0,2\pi) \} ,\; & \hbox{ if } z \in [\sigma/2,3\sigma/2],
           \\ 
            \{ (\theta, \phi)|\;\theta \in (0, \pi), \phi\in
            (0,2\pi)\} ,\;  &\hbox{ if } z \in
            [3\sigma/2,H-3\sigma/2] ,\\ 
            \{ (\theta, \phi)|\;\theta \in (\pi/2-b_2(z), \pi ),
            \phi\in(0, 2\pi)  \} ,\; & \hbox{ if } z \in [H-3\sigma/2,H-\sigma/2], \\
            \end{split}
    \end{aligned}
    \right.
\end{align} 
with
\begin{align} \label{ecu10}
\begin{aligned}
b_1(z) &= \sin ^{-1} \left( \frac{z- \sigma/2}{\sigma} \right), \\
b_2(z) &= \sin ^{-1} \left( \frac{H- \sigma/2- z}{\sigma} \right).
\end{aligned} 
\end{align}
In Fig. \ref{fig2} (color online), a schematic diagram of the possible angles of
collisions in a hard spheres system is plotted (purple region). It can be
appreciated that, depending on $z$, three parts in 
the system can be distinguished: one ``bulk'' part where all collisions are
possible and two ``boundary'' parts where the orientation of the
collisions is restricted due to one of the walls. The ``boundary''
part is the region between the dot line and the closest wall (grey
region). For $d=2$ the 
parametrization is similar. Defining $\theta$ as the angle between
$\widehat{\bm\sigma}$ and ${\widehat{ \bm e}}_{z }$, it is
\begin{align}\label{ecu9d2} \Omega_2(H, z) \equiv \left\lbrace 
\begin{aligned}
    \begin{split}
          \theta\in (-\pi/2-b_1(z), \pi/2+b_1(z)), \; &
           \hbox{ if } z \in [\sigma/2,3\sigma/2], 
           \\ 
           \theta \in (-\pi, \pi), \;  &\hbox{ if } z \in
            [3\sigma/2,H-3\sigma/2] ,\\ 
            \theta \in (\pi/2-b_2(z), 3\pi/2+b_2(z) ),\; &
            \hbox{ if } z \in [H-3\sigma/2,H-\sigma/2], \\ 
            \end{split}
    \end{aligned}
    \right.
\end{align} 
where $b_1$ and $b_2$ are defined in Eq. (\ref{ecu10}). 
Finally, the wall term is written as in \cite{dorfman2021contemporary}
\begin{equation}
L_W f({\bm r},{\bm v},t) =  \left[ \delta(z-H+\sigma/2)L_{\uparrow} +\delta(z-\sigma/2)L_{\downarrow}\right] f({\bm r},{\bm v},t),
\label{ecu11}
\end{equation}
with the operators given by
\begin{eqnarray}
  L_{\uparrow}  f({\bm r},{\bm v},t) &=&  \left[
  \Theta(-v_{z}-2v_p) |2v_p+v_z|b_{\uparrow}- \Theta(v_{z}) v_{z} \right] f(\bm{r}, \bm{v}, t)  ,
    \label{ecu12}\\
 L_{\downarrow}  f({\bm r},{\bm v},t) &=&  \left[
  \Theta(v_{z}-2v_p) |2v_p-v_z|b_{\downarrow}- \Theta(-v_{z}) |v_{z}| \right] f(\bm{r}, \bm{v}, t)  .
\label{ecu13}
\end{eqnarray}
Note that in the particle-particle collisional contribution defined in
Eq. (\ref{ecu6}), the particles are considered to be at two different
points, $\bm{r}$ and $\bm{r}_1$. This is in contrast with the
``traditional'' Boltzmann equation and it is necessary in order to
obtain a consistent equation. 

In the following, we will assume that the system is thin and dilute
enough so that 
the dependence of $f(\bm{r}, \bm{v}, t)$ on $z$ can be
neglected. Under this approximation the distribution function can be
substituted by its average along the $z$-coordinate, i.e.  
\begin{equation}
   f(\bm{r}, \bm{v}, t) \approx f(\bm{r}_{\parallel}, \bm{v}, t)
   \equiv \frac{1}{(H-\sigma)} \int_{\sigma/2}^{H-\sigma/2} \dd{z} ~
   f(\bm{r}, \bm{v}, t), 
\label{ecu14}
\end{equation}
with $\bm{r} \equiv
\bm{r}_{\parallel} + z \widehat{\bm e}_{z}$. Therefore, integrating
Eq.~\eqref{ecu5} over $z$, and taking into account the different
domains of solid angles defined in Eqs.~\eqref{ecu9}, gives rise to  
a closed evolution equation for $ f(\bm{r}_{\parallel}, \bm{v}, t)$ given by 
\begin{equation}
\left(
\frac{\partial}{\partial t} +
\bm{v}_{\parallel} \cdot \frac{\partial}{\partial \bm{r}_{\parallel}}\right)
f(\bm{r}_{\parallel}, \bm{v}, t)  
 =  \frac{1}{(H-\sigma)}\int_{\sigma/2}^{H-\sigma/2}dz\Jz
 +  \frac{1}{(H-\sigma)}( L_{\uparrow} + L_{\downarrow})
 f(\bm{r}_{\parallel}, \bm{v}, t). 
\label{ecu15}
\end{equation}
Interestingly, the particle-particle collision term verifies 
\begin{equation}\label{Jproperty}
\frac{1}{(H-\sigma)}\int_{\sigma/2}^{H-\sigma/2}dz\Jz=
\frac{\varepsilon-1}{\varepsilon} \mathcal{J}[f|f]
+\frac{1}{\varepsilon}\mathcal{\widetilde J}[f|f], 
\end{equation}
where we have defined
\begin{equation}
\J =  \sigma^{d-1}  \int \dd \bm{v}_{1} \int_{\Omega_{T,d}}\dd
\widehat{\bm\sigma}   \Theta(-\bm{g}\cdot\widehat{\bm\sigma})  
|\bm{g}\cdot\widehat{\bm \sigma}|(\alpha^{-2}b_{\sigma}^{-1} - 1) f(\bm{r}_{\parallel}, \bm{v}_1, t) f(\bm{r}_{\parallel}, \bm{v}, t),
\label{ecu16}
\end{equation}
and
\begin{equation}
\Jt =  \sigma^{d-2}\int^{3 \sigma/2}_{\sigma / 2} \dd z\int \dd
\bm{v}_{1} \int_{\Omega_d(2\sigma, z)} \dd \widehat{\bm\sigma} \,\Theta(-\bm{g}\cdot\widehat{\bm\sigma}) 
|\bm{g}\cdot\widehat{\bm \sigma}|(\alpha^{-2}b_{\sigma}^{-1} - 1) f(\bm{r}_{\parallel}, \bm{v}_1, t) f(\bm{r}_{\parallel}, \bm{v}, t),
\label{ecu17}
\end{equation}
with $\Omega_{T,d}$ being the total solid angle in $d$ dimensions, and
$\varepsilon \equiv (H-\sigma)/\sigma$. Hence, it splits in two
terms, $\frac{\varepsilon-1}{\varepsilon} \mathcal{J}$ and
$\varepsilon^{-1}\mathcal{\widetilde J}$. The first term corresponds
with the collisional contribution of a system of height
$(H-2\sigma)$ without  restrictions in the angular integration and the
second term is the collisional term of a system of height $2\sigma$
taking into account all the geometrical constraints. For $H\sim 
2\sigma$ ($\varepsilon\sim 1$), 
$\frac{\varepsilon-1}{\varepsilon}\mathcal{J}$ can be neglected 
compared to $\varepsilon^{-1}\mathcal{\widetilde J}$. In this case,
for $d=3$, 
Eq.~\eqref{ecu15} reduces to the one obtained in \cite{mgb19} for an
ultra-confined system with $H=2\sigma$. In the opposite
limit, i.e. $H\gg\sigma$($\varepsilon\gg 1$), the
$\frac{\varepsilon-1}{\varepsilon}\mathcal{J}$ terms dominates with 
respect to $\varepsilon^{-1}\mathcal{\widetilde J}$ and the dynamics
is the one of a ``bulk'' system of height $(H-2\sigma)\sim H$. This
is consistent with the fact that, in this limit, the geometrical
restrictions should not be relevant. Let us remark that, in order to
have this decomposition, the homogeneous in $z$ approximation given by
Eq. (\ref{ecu14}) is essential. In the same lines, the limit
  $H\gg\sigma$ is only meaningful if the above mentioned approximation is
  fulfilled. Henceforth, we will consider Eq.~\eqref{ecu15} as 
the dynamical equation of the system.

\section{Evolution equations for the horizontal and vertical temperatures}
\label{SecIII}

In the previous section, we have obtained a closed kinetic equation
for a confined system averaged over $z$ direction in the low density
regime, Eq.~\eqref{ecu15}, that admits a spatially homogeneous
solution of the form $f (\bm{v},t)$, i.e. a distribution independent
  of the horizontal spatial variables. In this section, the evolution
equations for the partial temperatures associated to the horizontal
and vertical degrees of 
freedom, $T$ and $T_z$, are studied for this spatially homogeneous
state. The partial temperatures $T$ and $T_z$ are defined as usual in
kinetic theory   
\begin{eqnarray}
(d-1)\frac{n}{2}    T(t) = \int \dd\bm{v} \, \frac{m}{2}   v_{\parallel}^2 f(\bm{v}, t),\\
\label{ecu18}
\frac{n}{2}    T_z(t) = \int \dd\bm{v} \, \frac{m}{2}   v_{z}^2 f(\bm{v}, t),
\label{ecu19}
\end{eqnarray}
with $v_{\parallel}^2 \equiv \sum_{i=1}^{d-1}v_{i}^2$  is the sum over all the horizontal 
coordinates and $n$ is the number density
defined by $n\equiv  N/(H-\sigma)L^{d-1}$. We assume that there is no
macroscopic velocity, i.e., $\int \dd {\bm v}\,\bm{v}   
f({\bm v},t) = \bm{0}$. This
assumption holds for all symmetric distribution functions in the variable ${\bm v}$, 
i.e., $f({\bm v}) = f({-\bm v})$. To obtain the evolution equations
for the temperatures, we take velocity moments in the Boltzmann-like
equation given by Eq.~\eqref{ecu15}. First, the evolution equation of  
the horizontal temperature is obtained by multiplying $\frac{m}{2}
v_{\parallel}^2 $ in Eq.~\eqref{ecu15} and integrating over all velocities. This leads to 
 \begin{equation}
 \label{ecu20}
 (d-1)\frac{n}{2}  \dv{T}{t} =
 \frac{m}{2(H-\sigma)}\int_{\sigma/2}^{H-\sigma/2}dz
\int \dd {\bm v}  \, v_{\parallel}^2 \, \Jz, 
 \end{equation}
where the wall collision term trivially vanishes because there is not energy 
injection in the horizontal direction. We proceed in the same way to
deduce the equation for vertical temperature. It can be written as
 \begin{equation}
 \label{ecu21}
 \frac{n}{2}  \dv{T_z}{t} = \frac{m}{2(H-\sigma)}\int_{\sigma/2}^{H-\sigma/2}dz
\int \dd {\bm v}  \, v_{z}^2 \, \Jz
 + \frac{m}{2(H-\sigma)} \int \dd {\bm v}  \,   v_{z}^2 ( L_{\uparrow}
 + L_{\downarrow}) f(\bm{v}, t). 
 \end{equation}
In this case, the wall terms appear since both walls inject energy in the vertical direction.  
To obtain closed equations for the partial temperatures, it is assumed that
the velocity distribution is an asymmetric maxwellian with two
different temperatures, associated to the horizontal and vertical directions
\begin{equation}
f(\bm{v}, t) = \frac{n}{\pi^{\frac{d}{2}} w^{d-1}(t)w_{z}(t)}  \,
{\rm exp}\bigg[-\frac{v^{2}_{\parallel} }{w^{2}(t)}-\frac{v^{2}_{z}}{w^{2}_{z}(t)} \bigg] , 
\label{ecu22}
\end{equation}
where we have introduced the thermal velocities
\begin{equation}
\frac{m}{2} w^{2}(t)  \equiv T(t), \label{ecu23}
\end{equation}
\begin{equation}
\frac{m}{2} w^{2}_{z}(t) \equiv  T_{z}(t). \label{ecu24}
\end{equation}
In the ansatz given by Eq. \eqref{ecu22} correlations between the
  horizontal and vertical degrees of freedom are neglected. This is
validated in the next sections by comparing the theoretical results
with MD simulations for $d=2$ and $d=3$. 

In order to solve the integrals given in Eqs. \eqref{ecu20} and \eqref{ecu21}, 
the following fundamental property of the particle-particle collision
term is used
\begin{align}\label{ecu25}
 \frac{1}{(H-\sigma)}\int_{\sigma / 2}^{H-\sigma / 2} \dd z \int \dd
  \bm{v} \, \psi(\bm{v})\Jz=& \frac{\sigma^{d-1}}{2(H-\sigma)}
                              \int_{\sigma / 2}^{H-\sigma / 2} \dd z
                              \int \dd \bm{v}_1 \int \dd \bm{v}
                              f(\bm{v}_1,t)f(\bm{v},t) \nonumber \\ 
 &   \times \int_{\Omega_d(H, z)} \dd \widehat{\bm{\sigma}}
   \;\abs{\bm{g} \cdot \widehat{\bm{\sigma}}} \Theta(\bm{g}
   \cdot \widehat{\bm{\sigma}} )  (   
 b_{{\widehat{\bm\sigma}}} -1) \{\psi(\bm{v}_1)+\psi(\bm{v}) \},
\end{align}
%introduced in Ref.~\cite{brilliantov2004kinetic}. 
where $\psi(\bm{v})$ is an arbitrary velocity function.
In Appendix \ref{Apendice_A1} the corresponding property for the
particle-wall collision operators is derived, obtaining
\begin{eqnarray}\label{ecu26}
 \int \dd \bm{v} \, \psi(\bm{v})L_{\uparrow} f(\bm{v},t) =&\int \dd \bm{v} \; f(\bm{v},t) \abs{v_z} \Theta(v_z)  (  
b_{\uparrow}-1) \psi(\bm{v}),\\\label{ecu27}
 \int \dd \bm{v} \, \psi(\bm{v})L_{\downarrow} f(\bm{v},t) =&\int \dd \bm{v} \; f(\bm{v},t) \abs{v_z} \Theta(- v_z)  (  
b_{\downarrow}-1) \psi(\bm{v}). 
\end{eqnarray}
Note that the left-hand side of Eqs.~\eqref{ecu25}-\eqref{ecu27}
corresponds to the rate of change of $\langle \psi \rangle \equiv \int
\dd v \, \psi f(\bm{v},t) $ due to collisions. In our case, we  study
the variation rate of temperature in each space coordinate, i.e., how
the mean values of $v_{\parallel}^2$ and $v_z^2$ change. Using  
the relations given in \eqref{ecu25}-\eqref{ecu27}, the collisional
integrals can be written as (see Appendix \ref{Apendice_B1})
\begin{align} \label{ecu28}
\begin{aligned}
\frac{m}{2(H-\sigma)}\int_{\sigma/2}^{H-\sigma/2}dz
\int \dd {\bm v}  \, v_{\parallel}^2 \, \Jz=&\frac{ m n^2 \sigma^{d-1}
  (1+\alpha)   \Omega_{T, d-1} \; w^3}
{  \sqrt{2\pi}\varepsilon } \int_0^1 \dd y \; \big( 1+ \beta y^2  \big)^{1/2} (1-y^2)^{\frac{(d-1)}{2}} \\ 
&\times      \left[\frac{(1+\alpha)}{2}\big( 1+ \beta y^2  \big)    -
  1 \right]
[(\varepsilon-1)+(1-y)],  \\
\end{aligned}
\end{align}
\begin{align} \label{ecu29}
\begin{aligned}
 \frac{m}{2(H-\sigma)}\int_{\sigma/2}^{H-\sigma/2}dz
\int \dd {\bm v}  \, v_{z}^2 \, \Jz=& \frac{ m n^2 \sigma^{d-1} (1+\alpha)   \Omega_{T, d-1}
  \; w^3}{  \sqrt{2\pi} \varepsilon} \int_0^1 \dd y \; \big( 1+ \beta y^2
 \big)^{1/2}    (1-y^2)^{\frac{(d-3)}{2}}  \\  
&\times   y^2 \left[\frac{(1+\alpha)}{2}\big( 1+ \beta y^2  \big)    -
  (1+\beta) \right]
[(\varepsilon-1)+(1-y)],
\end{aligned}
\end{align}
where $\beta \equiv \frac{T_z}{T}-1$. Note that both integrals split
into two terms: one term proportional to $(\varepsilon-1)$, plus
other term that corresponds to the results of Ref. \cite{mgb19} for $H=2\sigma$
($\varepsilon=1$), in agreement with the property given by
Eq. (\ref{Jproperty}). On the 
other hand, the integral corresponding to collision with walls is  
% \begin{equation} \label{ecu30}
% \frac{m}{2H} \int \dd \mathbf{v} ( L_{\uparrow} +L_{\downarrow}) f(\mathbf{v}, t)=2 \frac{m n v_{p} w_z }{H}  \bigg( \frac{v_p}{\sqrt{\pi}} +  \frac{w_{z}}{2} \bigg).
% \end{equation}
\begin{equation} \label{ecu30}
\frac{m}{2(H-\sigma)} \int \dd \bm{v} v_z^2( L_{\uparrow}
+L_{\downarrow}) f(\bm{v}, t)= 2 \frac{  n v_{p} T_z }{(H-\sigma)}  . 
\end{equation}
Note that Eq.~\eqref{ecu30} is twice the injection of energy term 
obtained in a system with a vibrated monolayer 
when only one  wall is moving \cite{mgb19}. This equivalence is
consistent because the double of energy is injected by the walls. 

The integrals \eqref{ecu28} and \eqref{ecu29} can be solved exactly,
but the expressions are quite complex to handle with. To get a simpler
theoretical analysis, we perform a first order expansion  considering
$\beta \ll 1$, i.e., the horizontal and vertical temperatures are
close. In this case, the equations are 
\begin{align}
\dv{T}{t} =&   \frac{ n \sigma^{d-1} (1 + \alpha)   \Omega_{T, d} \;
             w }{   \sqrt{2  \pi }  d\left(  2 + d \right) \varepsilon}
             T \left\lbrace (\varepsilon-1 ) \bigg[   - (2+d)(1 - \alpha)  
+ \frac{ (1 +3 \alpha)}{2}  \beta \bigg]   \right.   \label{ecu31} \\ \nonumber
& \left. - g(d)(1 - \alpha) 
+ \frac{ (1 +3 \alpha)h(d)}{2}  \beta\right\rbrace
 ,
\\
 \dv{T_z}{t}  =& - \frac{ n \sigma^{d-1} (1 + \alpha)   \Omega_{T, d}
                 \;  w }{   \sqrt{2  \pi }  d\left(  2 + d \right)
                 \varepsilon }     T \left\lbrace (\varepsilon-1)\bigg[
                 (2+d)(1 - \alpha)  
+ \frac{ (5+4d - 9 \alpha)}{2}  \beta \bigg]   \right. \label{ecu32} \\ \nonumber
& \left.  g_{z}(d)(1 - \alpha) 
+ \frac{4g_{z}(d) - (1 +3 \alpha)h_{z}(d)}{2}  \beta \right\rbrace 
+  \frac{ 4   v_{p} T_z }{\sigma\varepsilon} ,
\end{align}
with 
\begin{eqnarray} 
        g (d ) & \equiv&  (2+d) -  \frac{4 (3+d)
                 \gammaf{2+\frac{d}{2}}}
{\left[ (2+d)^2 - 1\right]  \sqrt{\pi}  \gammaf{\frac{d+1}{2}}}  ,  \label{ecu33} \\ 
        h (d ) & \equiv& 1 -  \frac{8 \gammaf{2+\frac{d}{2}}}
{ \left[ (2+d)^2 - 1\right]\sqrt{\pi}  \gammaf{\frac{d+1}{2}} }, \label{ecu34}   \\
       g_{z} (d ) & \equiv& (2+d)  -\frac{16  \gammaf{2+\frac{d}{2}} }{ (d^2-1) \sqrt{\pi}  \gammaf{\frac{d-1}{2}}} \label{ecu35}  ,   \\
        h_{z} (d ) & \equiv &
         6  -\frac{64 \gammaf{2+\frac{d}{2}}}{(3+d)(d^2-1)\sqrt{\pi}
                     \gammaf{\frac{d-1}{2}}},   \label{ecu36} 
\end{eqnarray}
where $\gammaf{z} \equiv \int_0^{\infty}\dd t \;   t^{z-1} e^{-t}$ is
the Gamma function. Let us briefly analyze Eqs.~\eqref{ecu31} and \eqref{ecu32}. First,
note that, in agreement with Eq. (\ref{Jproperty}), two separated
terms are obtained in the particle-particle collisional contribution
of both equations:  One of them
is multiplied by a factor $(\varepsilon-1)$ becoming dominant for $H \gg
\sigma$, while the other  dominates for $H=2 \sigma$, where the first
term vanishes. If the zeroth order in $\beta$ is
considered and $v_p=0$ is taken, in the limit of $H\gg\sigma$, the
temperature evolution equation of a freely evolving granular  
gas is obtained in the gaussian approximation \cite{van1998velocity}. We also 
remark that the Eqs.~\eqref{ecu31} and \eqref{ecu32} are obtained
assuming $\beta\ll 1$, so that the equations are only valid 
if the horizontal and vertical temperatures are of the same  order of
magnitude, i.e., $T \approx T_z$. The $\beta$ terms in
Eqs.~\eqref{ecu31} and \eqref{ecu32} describe the energy transfer
between the vertical and horizontal degrees of freedom due to particle
collisions, while the last term of Eq.~\eqref{ecu32} describes the
energy injection term due to the walls. Thus, the dynamics can be
summarized as follows: wall-particle collisions inject energy in the
vertical direction,  particle-particle collisions transfer energy from
the vertical to the horizontal degrees of freedom, while energy is
dissipated through inelastic collisions. The main difference with the
corresponding equations in the ultra-confined system \cite{mgb19}, is
that energy is also dissipated in the vertical degree of freedom. 

From Eqs. (\ref{ecu31}) and (\ref{ecu32}) the stationary partial
temperatures, $T_s$ and $T_{z, s}$, can be obtained. In particular,
from Eq. (\ref{ecu31}), the quotient between the stationary
temperatures, $\beta_s\equiv\frac{T_{z, s}}{T_s}-1$, is easily calculated, obtaining
\begin{equation}\label{betaST}
\beta_s=K(d, \varepsilon)\frac{1-\alpha}
{1+3\alpha}, 
\end{equation}
where 
\begin{equation}
K(d,\varepsilon)=2\frac{(d+2)(\varepsilon-1)+g(d)}
{\varepsilon-1+h(d)}, 
\end{equation}
is a geometric factor that depends on the height of the system and the
spatial dimensionality. 
Hence, $\beta_s$ depends on the inelasticity, height and $d$, but it
is density independent. 
$\beta_s$ is always positive in the range $0<\alpha<1$, so that
$T_{z, s}$ is always larger than $T_s$, consistent with the fact that
energy is injected in the vertical direction and then transferred to
the horizontal one. Moreover, for a given height, $\beta_s$ decreases
monotonically with the inelasticity, vanishing in the elastic limit, in
accordance with equipartition. Note that this property is not obvious, 
as the elastic limit could be singular (equipartition obviously holds in
the elastic case with $v_p=0$). At constant inelasticity, $\beta_s$
decays monotonically with the height. Intuitively, this was expected to
happen because, the thinner the 
system is, the more anisotropic is. In the $\varepsilon\gg 1$
  limit, the following simple asymptotic expression is obtained 
\begin{equation}
\beta_s\sim\beta_s^{(\infty)}\equiv\frac{2(d+2)(1-\alpha)}{1+3\alpha},\quad\text{for}\quad
\varepsilon\gg 1. 
\end{equation}

\begin{figure}[t]
 \centering \includegraphics[scale=0.34]{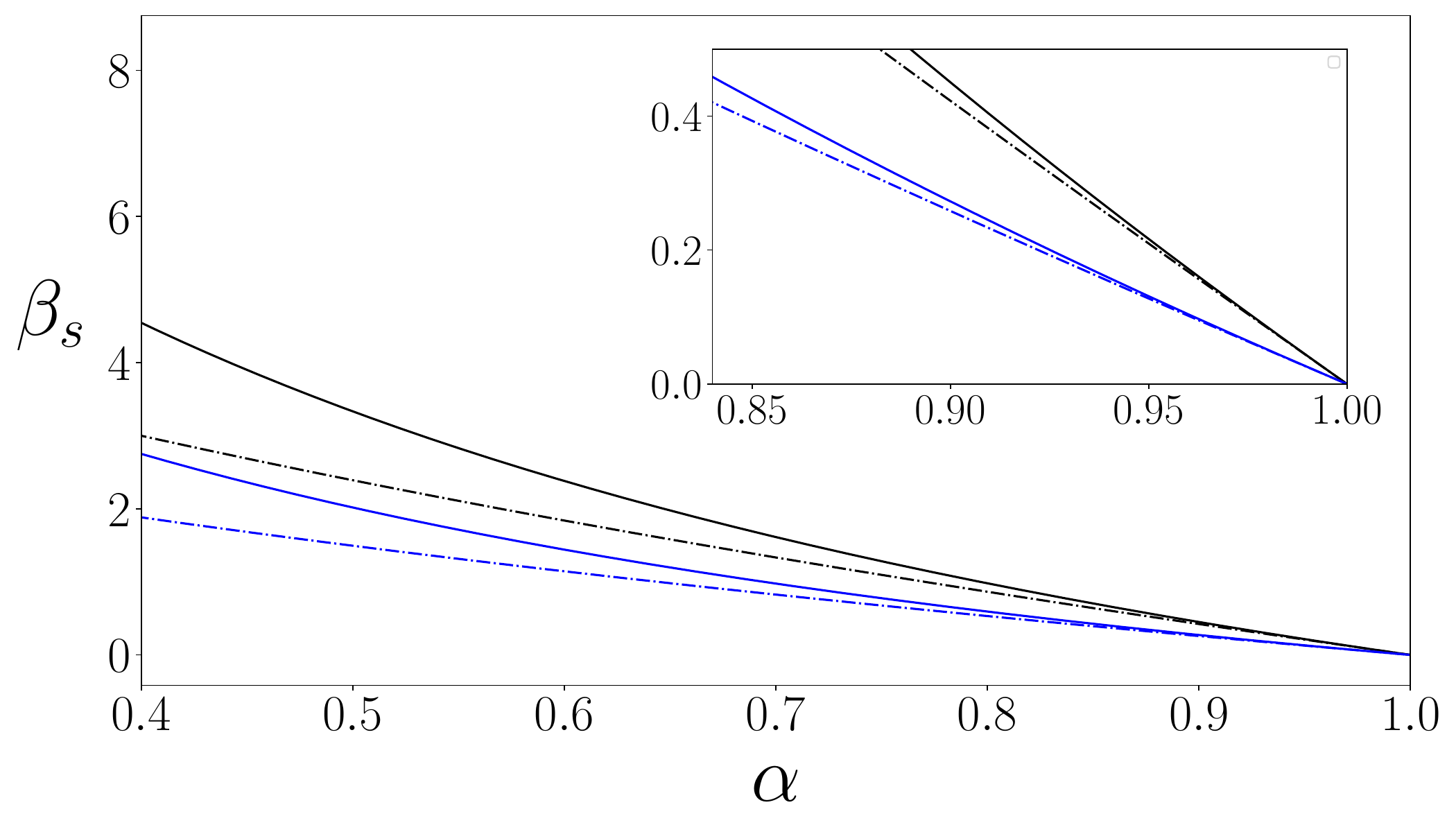}
 \caption{(Color online) Theoretical expressions for $\beta_s$
given by Eq. (\ref{betaST}) (solid lines) and by the numerical
solution coming from
(\ref{ecu28}) (dashed lines) for $d=3$ as a function of
the inelasticity. Two values of the height are considered: 
$H=2\sigma$ (black) and $H=29\sigma$ (blue). The
quasi-elastic region is enlarged in the inset. }
\label{beta_alfa_teo}
\end{figure}

\begin{figure}[t]
 \centering \includegraphics[scale=0.34]{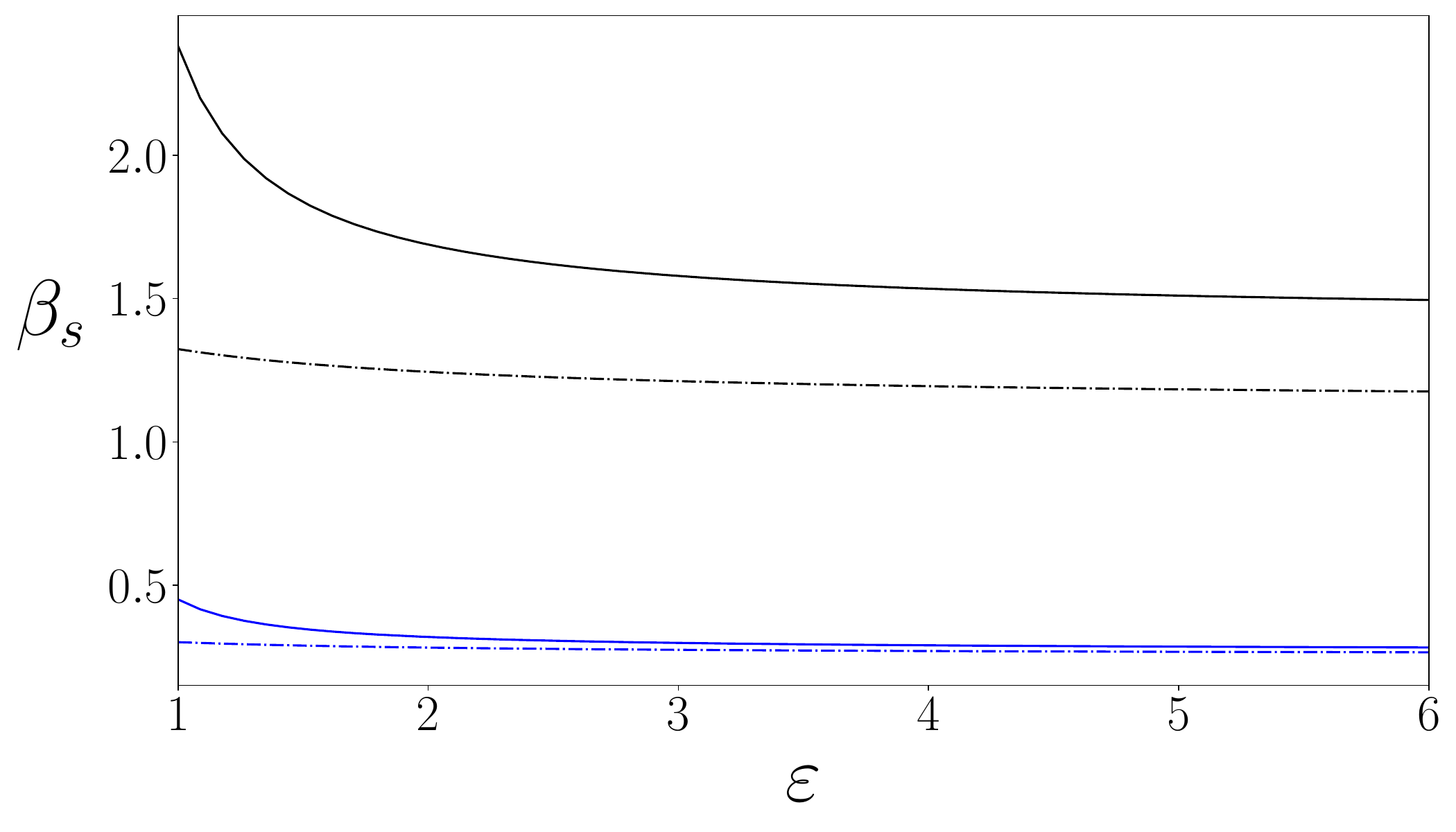}
 \caption{(Color online) Theoretical expressions for $\beta_s$
given by Eq. (\ref{betaST}) (solid lines) and by the numerical
solution coming from
(\ref{ecu28}) (dashed lines) for $d=3$ as a function of
the dimensionless height, $\epsilon$. Two values of the inelasticity are considered: 
$\alpha=0.6$ (black) and $\alpha=0.9$ (blue). }
\label{beta_h_teo}
\end{figure}

The expression for $\beta_s$ given
by Eq. (\ref{betaST}) is only valid in the small $\beta$ case. Nevertheless,
from Eq. (\ref{ecu28}) a closed equation for $\beta_s$ can be obtained
that is valid, in principle, for arbitrary $\beta$ (of course, its
validity depends on the validity of the maxwellian ansatz) and that
has been solved numerically. 
%The
%transcendental equation for $\beta_s$ has been solved by applying the
%\textit{Powell's dog leg method} \cite{powell1970hybrid} given in
%\texttt{Scipy} Python package \cite{2020SciPy-NMeth}. 
In Fig. \ref{beta_alfa_teo} (color online) the approximate expression for $\beta_s$
given by Eq. (\ref{betaST}) (solid lines) and the one coming from Eq. 
(\ref{ecu28}) (dashed lines) are plotted for $d=3$ as a function of
the inelasticity. We have considered two values of the height,
$H=2\sigma$ (black) and $H=29\sigma$ (blue). The
quasi-elastic region is enlarged in the inset. It can be
seen that the approximate expression for $\beta_s$ and the one coming from
Eq. (\ref{ecu28}) are very similar from the elastic limit till
$\alpha\sim 0.8$ (where $\beta\sim 1$). Similar results are obtained
for $d=2$. In Fig. \ref{beta_h_teo} (color online) the approximate
expression for $\beta_s$ given by Eq. (\ref{betaST}) (solid lines) and
the one coming from Eq.  
(\ref{ecu28}) (dashed lines) are plotted for $d=3$ as a function of
the dimensionless height, $\epsilon$. We have considered two values of
the inelasticity, 
$\alpha=0.6$ (black) and $\alpha=0.9$ (blue). For $\alpha=0.9$ both
curves are very similar except close to $\varepsilon=1$ ($H=2\sigma$)
where $\beta_s$ is not sufficiently small for the approximation to be
valid. For $\alpha=0.6$, the inelasticity induces a stronger
anisotropy \cite{bgm19} in such a way that the approximation is no longer accurate,
independently of the height of the system. For both inelasticities it
is seen that the bulk expression accurately describes a wide range of
heights, for values of $\varepsilon$ larger than approximately $2$. 

%\begin{figure}[t]
% \centering \includegraphics[scale=0.34]{Figures/betas_vs_alpha_withoutinst_tendence.pdf} 
% \caption{ Values of $\beta_s$ as a function of $\alpha$ for $d=2$. Eq.~\eqref{ecu51} is represented for 
% $H=2\sigma$ (black solid line) and $H=29\sigma$ (red dashed line).  The inset displays the limit $\alpha \to 1 $ for both heights.   }
%\label{beta_alfa_teo_d2}
%\end{figure}

By Inserting $\beta_s$ giving by Eq. (\ref{betaST}) into
Eq. (\ref{ecu32}), the stationary horizontal temperature can be
calculated, obtaining 
\begin{equation}\label{TsTheory}
\left[\frac{T_s}{mv_p^2}\right]^{1/2}=\frac{4\sqrt{\pi}d(d+2)(1+\beta_s)}
{(1+\alpha)\Omega_{T, d}\left[C(d, \alpha, \varepsilon)
+D(d, \alpha,\varepsilon)(\varepsilon-1)]\right\}n\sigma^d}, 
\end{equation}
where the following functions 
\begin{eqnarray}
C(d, \alpha,\varepsilon)&=&
(1-\alpha)g_z(d)+[4g_z(d)-(1+3\alpha)h_z(d)]\frac{\beta_s}{2},\\
D(d, \alpha,\varepsilon)&=&
(1-\alpha)(d+2)+(5+4d-9\alpha)\frac{\beta_s}{2}, 
\end{eqnarray}
have been introduced. Independently of the complex behavior of $T_s$
as a function of the inelasticity and the height, that will be
discussed later, the density
dependence is very simple, $T_s\sim (n\sigma^d)^{-2}$. This dependence
comes from the fact that the number of wall-particle collisions grows
with $n\sigma^d$, while the number of particle-particle collisions grows
with $(n\sigma^d)^2$. Then, the more dilute the system is, the more
energy is injected. Moreover, the stationary pressure defined as $p
(n) \equiv n T_s (n)$ goes as $(n\sigma^d)^{-1}$ and the steady state
compressibility, $\kappa \equiv\dv{p}{n} $, is negative. 
This macroscopic property is similar to the one observed in vibrated
ultra-confined Q2D systems \cite{mgb19b}, in which the
negative compressibility is the essential ingredient that triggers the
instability. We will further discuss this point in the last 
section. Note that the temperature diverges in  
the elastic limit, $\alpha \to 1 $, because there is no energy loss
mechanisms. For $\varepsilon\gg 1$, Eq. (\ref{TsTheory}) is
  simplified, obtaining 
\begin{equation}
\left[\frac{T_s}{mv_p^2}\right]^{1/2}\sim\frac{4\sqrt{\pi}d(d+2)(1+\beta_s^{(\infty)})}
{(1+\alpha)\Omega_{T, d}D(d, \alpha,\infty)\varepsilon n\sigma^d}, \quad\text{for}\quad
\varepsilon\gg 1, 
\end{equation}
where $D(d, \alpha,\infty)\equiv\lim_{\varepsilon\to\infty}D(d,
\alpha,\varepsilon)
=(1-\alpha)(d+2)+(5+4d-9\alpha)\frac{\beta_s^{(\infty)}}{2}$. Note
that, in this asymptotic regime, the dependence on the size of the system
is very simple, i.e. $T_s\propto\varepsilon^{-2}$. 
On the
other hand, we have also   
obtained $T_s$ using the exact solutions given in Eqs.~\eqref{ecu27}
and \eqref{ecu28}. However, no difference between both
expressions are appreciated in the relevant range of parameters.  

The objective now it to study the stability of the stationary solution
in the context of Eqs.~\eqref{ecu31} and
\eqref{ecu32}, i.e., it is assumed that there are not gradients and
that the dynamics is given in terms of the above mentioned
equations. In order to perform the analysis, it is convenient to
introduce the following dimensionless time scale
\begin{align} \label{ecu37}
s(t) =\frac{ n \sigma^{d-1} (1 + \alpha)   \Omega_{T, d} \ }{
  \sqrt{2  \pi }  d\left(  2 + d \right) \varepsilon }  \int^{t}_{0}d\tau
  w(\tau), 
\end{align}
that is proportional to the collisions per particle
in $(0,t)$. We will consider small deviations around the
stationary temperatures, so that  Eqs.~\eqref{ecu31} and
\eqref{ecu32} can be linearized. 
%Considering the above definition, the evolution equations become
%\begin{align}
%\dv{\widetilde T}{s} =&       \widetilde{T} \left\lbrace - (1 -
%\alpha)  \bigg[    (2+d) \gamma 
% +  \frac{g(d)}{\varepsilon}   \bigg]   +   \bigg[   \gamma 
%+ \frac{ h(d)}{\varepsilon} \bigg]  \frac{\beta(1+3\alpha)}{2}  \right\rbrace \label{ecu38}
% ,
%\\
% \dv{\widetilde{T}_z}{s}  =& -     \widetilde{T} \left\lbrace   (1 -
% \alpha)  \bigg[    (2+d) \gamma 
% +  \frac{g_{z}(d)}{\varepsilon}    \bigg]   +  \bigg[  (5+4d-9\alpha) \gamma 
%+ \frac{(4g_{z}(d) - (1 +3 \alpha)h_{z}(d) )}{ \varepsilon}\bigg] \frac{\beta}{2}\right\rbrace %  \label{ecu39} \\ \nonumber
%& + \frac{ 4   \sqrt{ \pi }  \left(  2 + d \right) d}  { n
%  \sigma^{d-1}(H-\sigma) (1 + \alpha)   \Omega_{T, d} }
%\frac{\widetilde{T}_z }{\sqrt{\widetilde T}} , 
%\end{align}
%where dimensionless temperatures have been introduced
%\begin{align}
%\widetilde{T} & \equiv \frac{T}{mv_p^2}, \label{ecu40} \\
%\widetilde{T}_z & \equiv \frac{T_z}{mv_p^2} \label{ecu41}. 
%\end{align} 
The deviations $\delta T \equiv T -T_s$ and $\delta T_{z} \equiv
T_{z}-T_{z, s}$ obey the following system of ordinary differential  
equations that can be expressed in matrix form as follows
\begin{equation} \label{ecu42}
\frac{d}{d s}\begin{pmatrix}
\delta T \\
\delta T_{z}
\end{pmatrix}=M\begin{pmatrix}
\delta T \\
\delta T_{z}
\end{pmatrix},
\end{equation}
where $M$ is the matrix
\begin{align} \label{ecu43}
    M = -\begin{pmatrix} 
    (1 - \alpha) A   + \frac{B}{2} & \; -  \frac{B}{2} \\
 \frac{3(1-\alpha)}{2} A_z+ \frac{(\beta_s-2)}{2} \frac{B_z}{2} & \frac{1}{1+\beta_s}\Big[ \frac{B_z}{2}- (1-\alpha)A_z \Big] 
    \end{pmatrix},
\end{align}
with
\begin{align}
    A \equiv&      (2+d)(\varepsilon-1)
 +  g(d)   , \label{ecu44}\\
   B \equiv&  [\varepsilon-1+h(d)](1+3\alpha), \label{ecu45}\\
  A_z \equiv&   (2+d)(\varepsilon-1)
 +  g_{z}(d)  , \label{ecu46}\\
   B_z \equiv&    (5+4d-9\alpha) (\varepsilon-1)
+ 4g_{z}(d) - (1 +3 \alpha)h_{z}(d)  . \label{ecu47}
\end{align}
%\begin{figure}[t]
% \centering \includegraphics[scale=0.30]{Figures/eigen3d.pdf}
% \caption{ Real part (solid line) and imaginar part (dashed line) for
 %  Eigenvalues $\lambda_i$, with $i=1,2$, associated with $M$ matrix
  % \eqref{ecu43} as a function of inelasticity coefficient. The
  % eigenvalues for $H=2\sigma$ (black line) and $H=29\sigma$ are
   %represented.} 
%\label{fig_eigen_3d}
%\end{figure}
The eigenvalues of the matrix $M$ have been calculated numerically obtaining that 
its real part is always negative, so that the stationary state is
linearly stable. For mild inelasticities, two different real eigenvalues are
obtained, $\lambda_2<\lambda_1<0$, so that, for long times, the
dynamics is dominated by the slowest eigenvalue,
$\lambda_1$, that defines the \emph{hydrodynamic} time
scale. Interestingly, for strong inelasticities, the eigenvalues turn
complex, and the stationary state is reached oscillating. This happens
for $\alpha\sim 0.7$, independently of the other parameters. The
origin and the possible implications in the hydrodynamic description
of the system 
of this transition will be studied in detail elsewhere \cite{mgmTBP}.

\section{Simulation results}
\label{SecIV}

\begin{figure}[t!]
  \centering \includegraphics[scale=0.20]{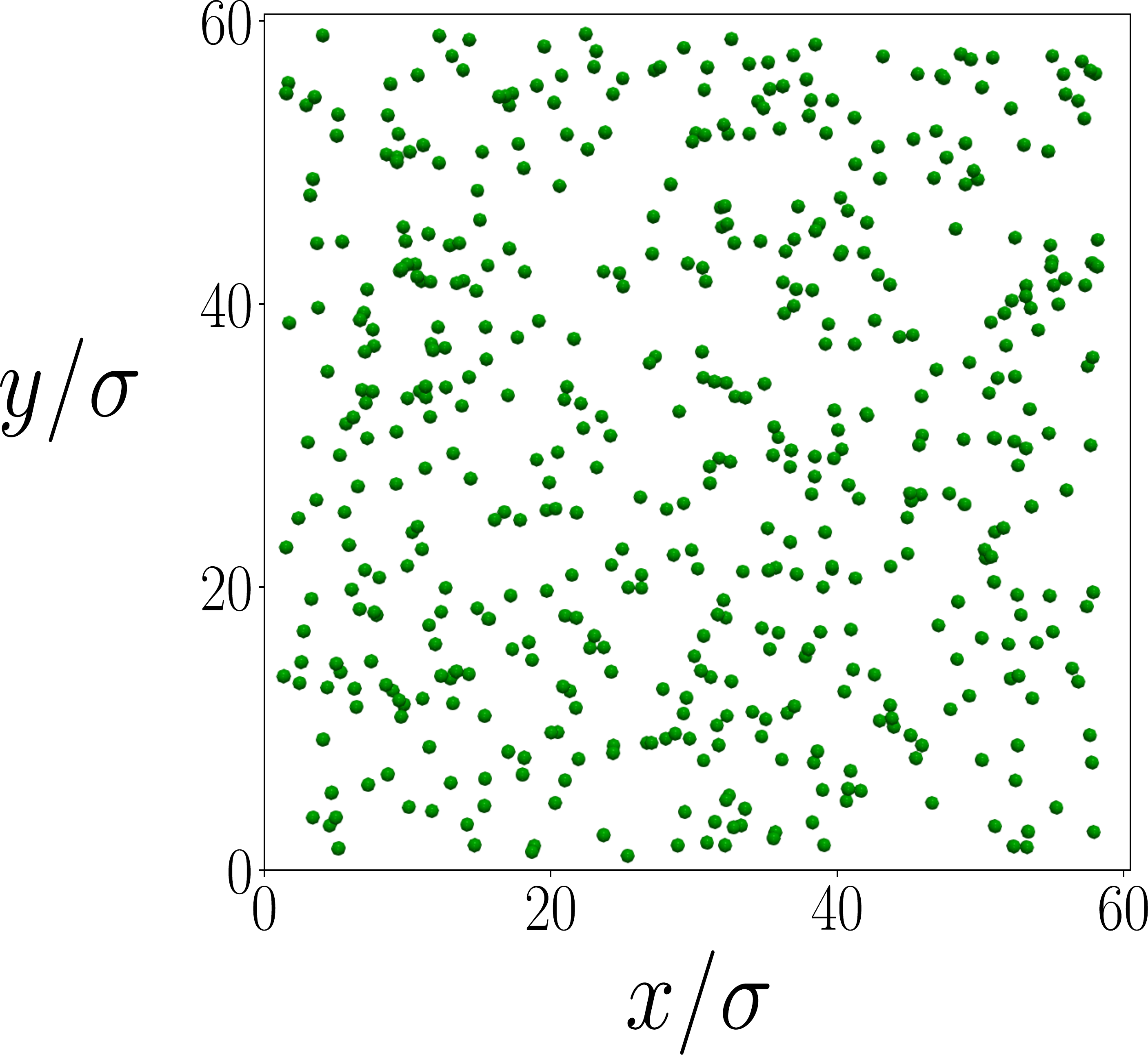}
 \caption{Snapshot of a typical configuration of a hard spheres system
   (observed from above). The values of the parameters
are the ones for $d=3$ with $H=8\sigma$ and $\alpha=0.9$.  } 
\label{figSnaphot}
\end{figure}

\begin{figure}[t!]
  \centering \includegraphics[scale=0.3]{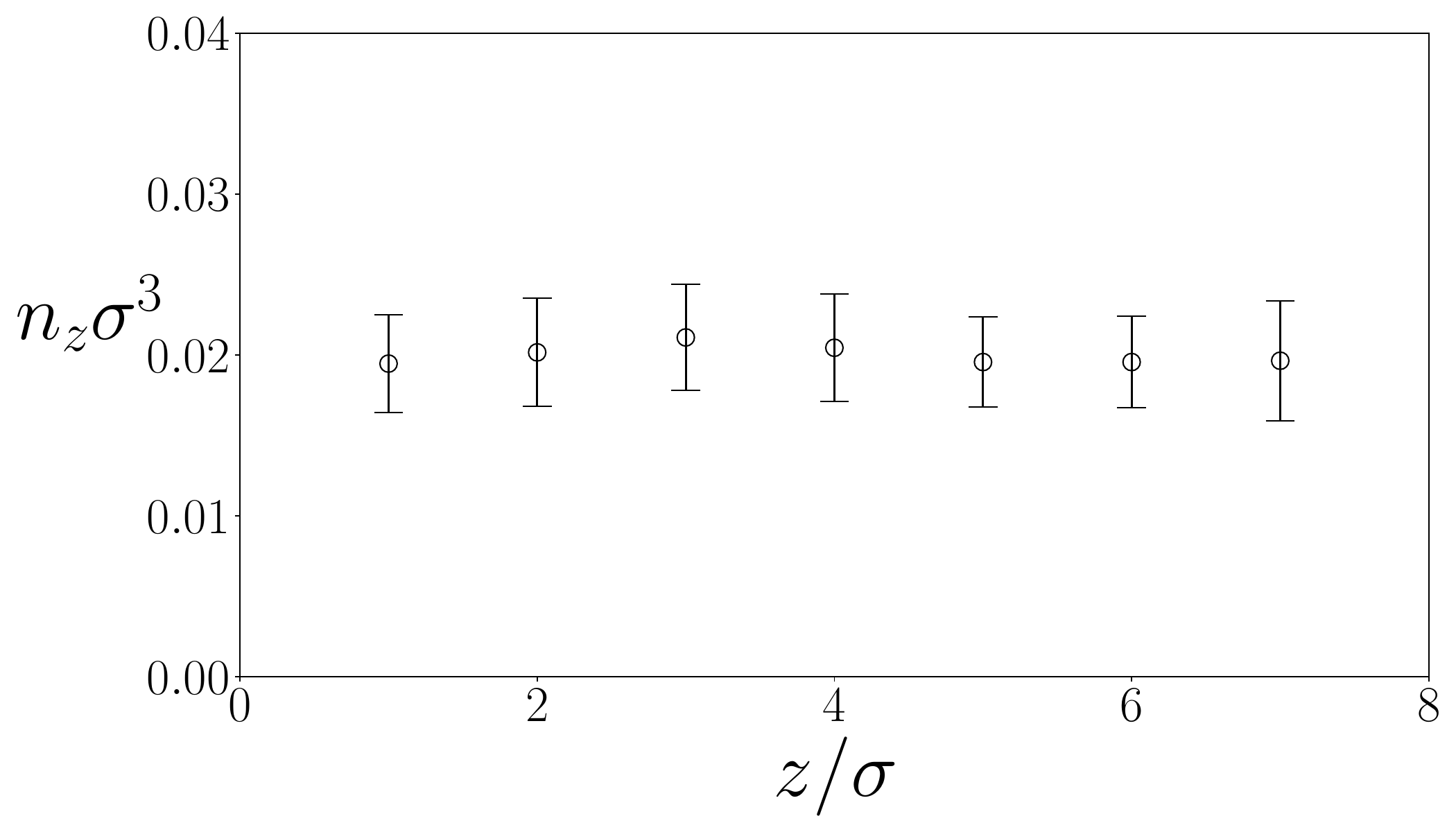}
 \caption{Dimensionless vertical averaged density, 
$n_z(z)\sigma^3$, as a function of
  $z/\sigma$ for a hard spheres system with $H=8\sigma$ and
  $\alpha=0.9$. The error bars are obtained from the average
over the $10$ realizations. } 
\label{figDens}
\end{figure}

\begin{figure}[t!]
  \centering \includegraphics[scale=0.4]{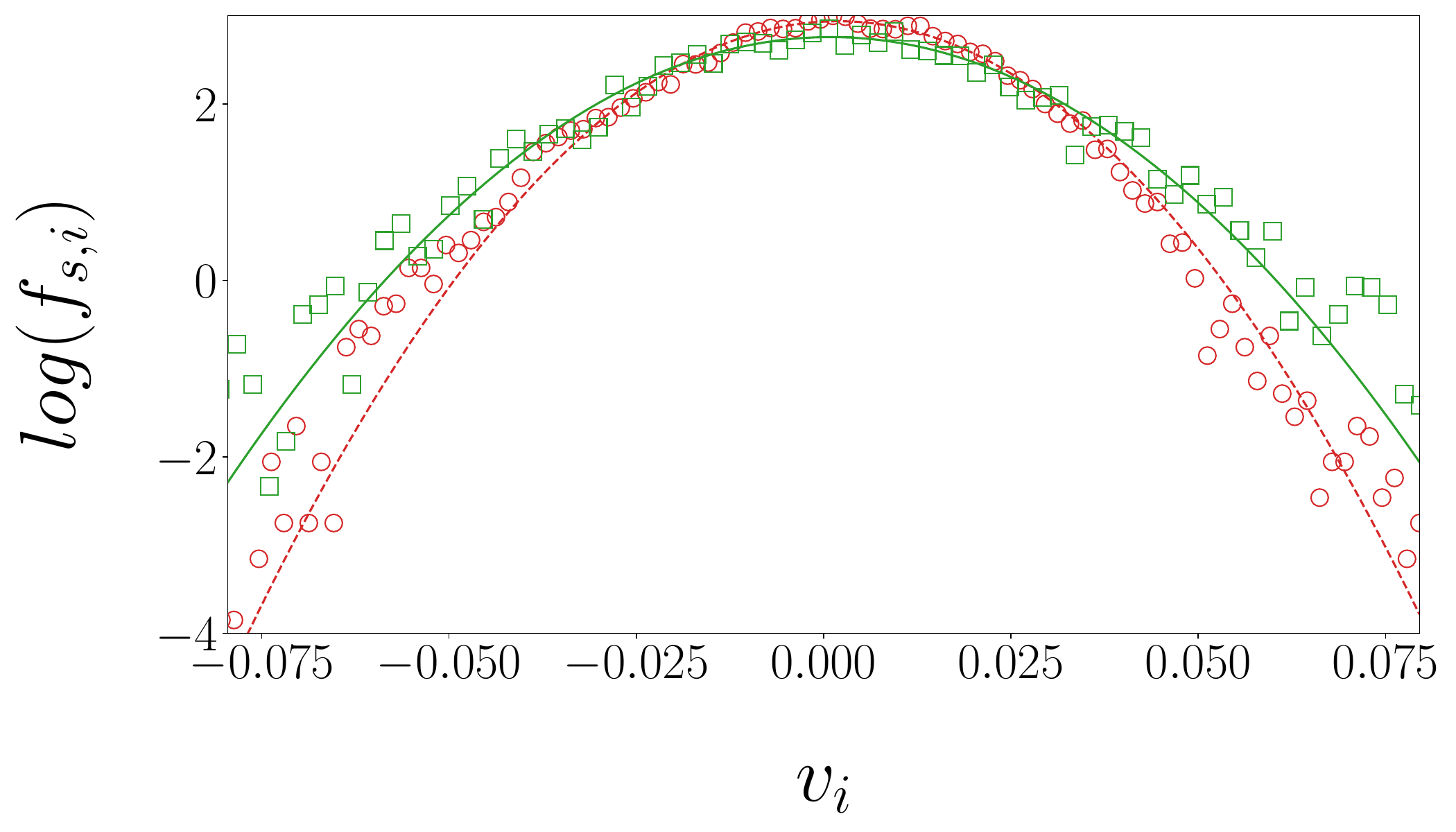}
 \caption{ Simulation results for the logarithm of the
stationary marginals velocity distribution functions, 
$f_{s,x}$ (red circles)
and $f_{s,z}$ (green
squares), as functions of $v_x$ and $v_z$, respectively, for
$H=5\sigma$ and $\alpha=0.65$. The (red) 
dashed line and the (green) solid line are the corresponding quadratic
interpolations.   }
\label{figPdf}
\end{figure}

By using the Event-Driven algorithm \cite{allen}, we have performed MD
simulations of an ensemble of inelastic hard spheres 
or disks confined between two sawtooth walls perpendicular to the
$z$-direction. The vibrating walls are separated a distance $H$ and
periodic boundary 
conditions are applied in the horizontal direction. We have taken $m$
and $\sigma$ as the unit of mass and length, respectively.  To
generate the initial condition, the particles are placed inside the
system at random, with a two-temperature gaussian
velocity distribution, being the initial horizontal temperature, $T(0)$, the
unit of energy. The initial vertical temperature has been chosen
between $0.1T(0)$ and $10T(0)$, depending on the simulation. The
velocity of both walls is 
$v_p=0.001\left[\frac{T(0)}{m}\right]^{1/2}$. For $d=3$, we have taken
$N=512$ and $n\sigma^3=0.02$ and for $d=2$, $N=300$ and
$n\sigma^2=0.02$, so that the system can be considered to be dilute
and the kinetic theory description of the previous section is expected
to be valid. The results have been averaged over $10$
realizations. Let us remark that the simulations in our confined
  system run much more slowly than in the non-confined system due to
  the huge number of collisions with the walls. In any case, the
  considered number of particles is large enough for the
  Boltzmann equation to be valid \cite{mgb19} and the number of
  realizations suffices to have a good statistic to validate the theory. For 
different values of the height and the inelasticity, 
we have checked that the system remains in a spatially homogeneous
state for all the time evolution. In Fig. \ref{figSnaphot} a typical
snapshot of the system (observed from above) is shown. No density
gradients are appreciated. The values of the parameters
are the ones for $d=3$ with $H=8\sigma$ and $\alpha=0.9$. 
We have
performed a more quantitative 
analysis by measuring the hydrodynamic fields in the horizontal and
vertical directions. For all the values of the parameters that are
considered in this section, it was found that the system is spatially homogeneous
with a high degree of accuracy. For example, the dimensionless
  vertical averaged density,  
$n_z(z)\sigma^3\equiv\frac{1}{L^2}\int dx\int dy\,
  n(\mathbf{r}) \sigma^3$, is plotted in Fig. \ref{figDens} as a function of
  $z/\sigma$ for a hard spheres system with $H=8\sigma$ and
  $\alpha=0.9$. The error bars are obtained from the average
over the $10$ realizations. The density fluctuates around its mean value,
  $n\sigma^3=0.02$, showing that the system is homogeneous in the
  vertical direction with a high degree of accuracy. Similar results
  are obtained for the other values of the parameters considered in this
section. 
For strong inelasticities, some
simulations showed the development of clusters of particles. Of
course, these simulations were discarded. 
We have also measured the marginals one-particle distribution
functions. In Fig. \ref{figPdf}
(color online), the simulation results for the logarithm of the
marginals velocity distribution functions once the stationary state is reached, 
$f_{s,x}(v_x)\equiv\int dv_y\int dv_zf_s(\mathbf{v})$ (red circles)
and $f_{s,z}(v_z)\equiv\int dv_x\int dv_yf_s(\mathbf{v})$ (green
squares), are plotted for $H=5\sigma$ and $\alpha=0.65$. The (red)
dashed line and the (green) solid line are the corresponding quadratic
interpolations that fit accurately the simulation results. Similar
results are obtained for the transient to the stationary state and for
other values of the parameters, showing the accuracy of the ansatz
given by Eq. (\ref{ecu22}). 

\begin{figure}[t]
\begin{center}
\begin{subfigure}{0.8\textwidth}
    \centering \includegraphics[scale=0.28]{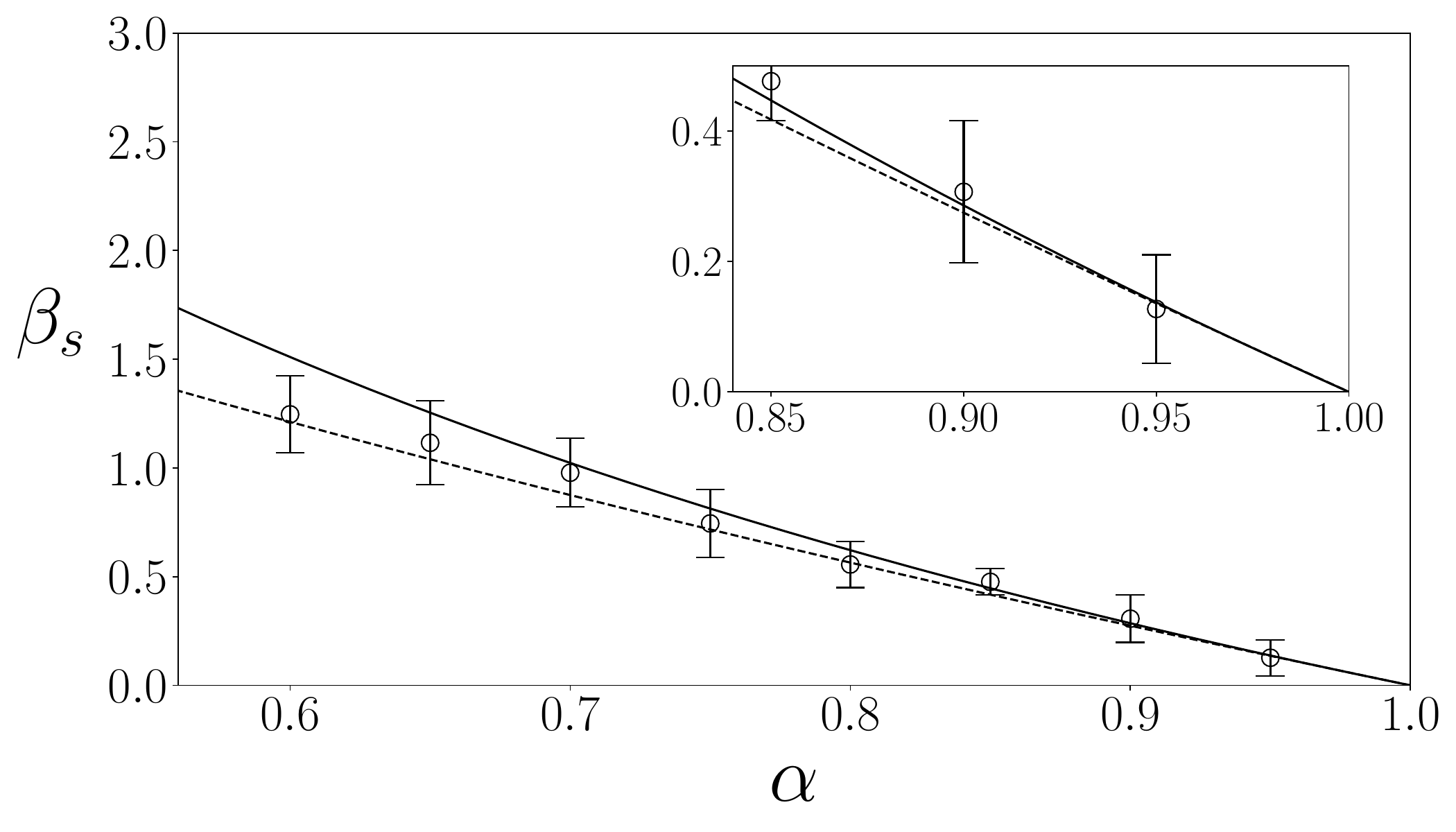}
    \caption{}
    \label{betaSH5}
\end{subfigure}
\hfill
\begin{subfigure}{0.8\textwidth}
 \centering \includegraphics[scale=0.28]{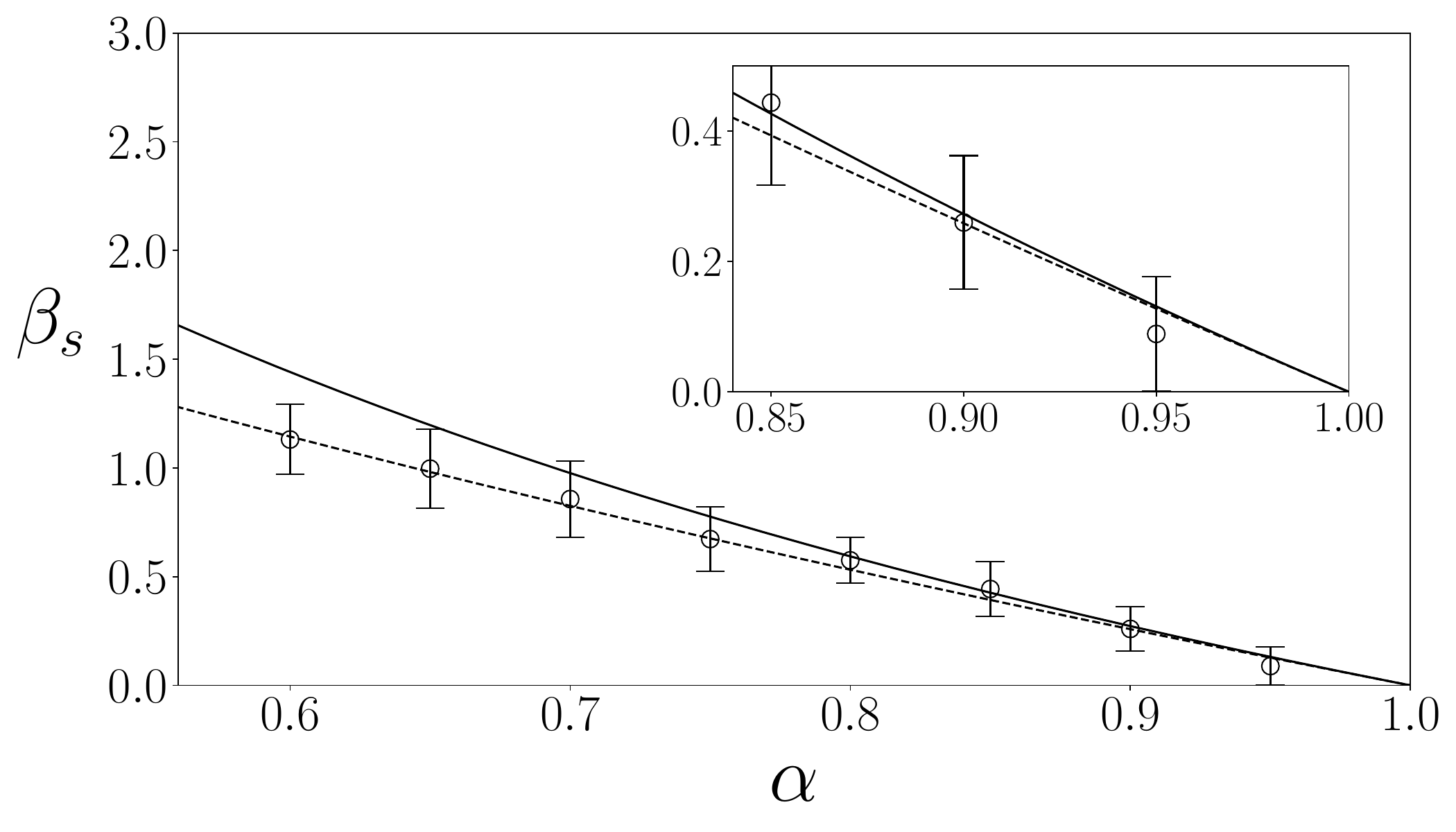}   
 \caption{}
 \label{betaSH29}
\end{subfigure}
\end{center}
\caption{$\beta_{s}$ of a hard spheres system as a function of $\alpha$
  for $H=5\sigma$ (a) and $H=29\sigma$ (b). The circles are the
  simulation results, the solid line 
is the linear approximation of $\beta_s$ given by Eq. (\ref{betaST})
and the dashed line is the numerical theoretical prediction coming
from Eq. (\ref{ecu28}). The error bars are obtained from the average
over the $10$ realizations. The quasi-elastic region is enlarged in the
inset.  }
\label{beta_vs_alpha}
\end{figure}

\begin{figure}[t!]
  \centering \includegraphics[scale=0.3]{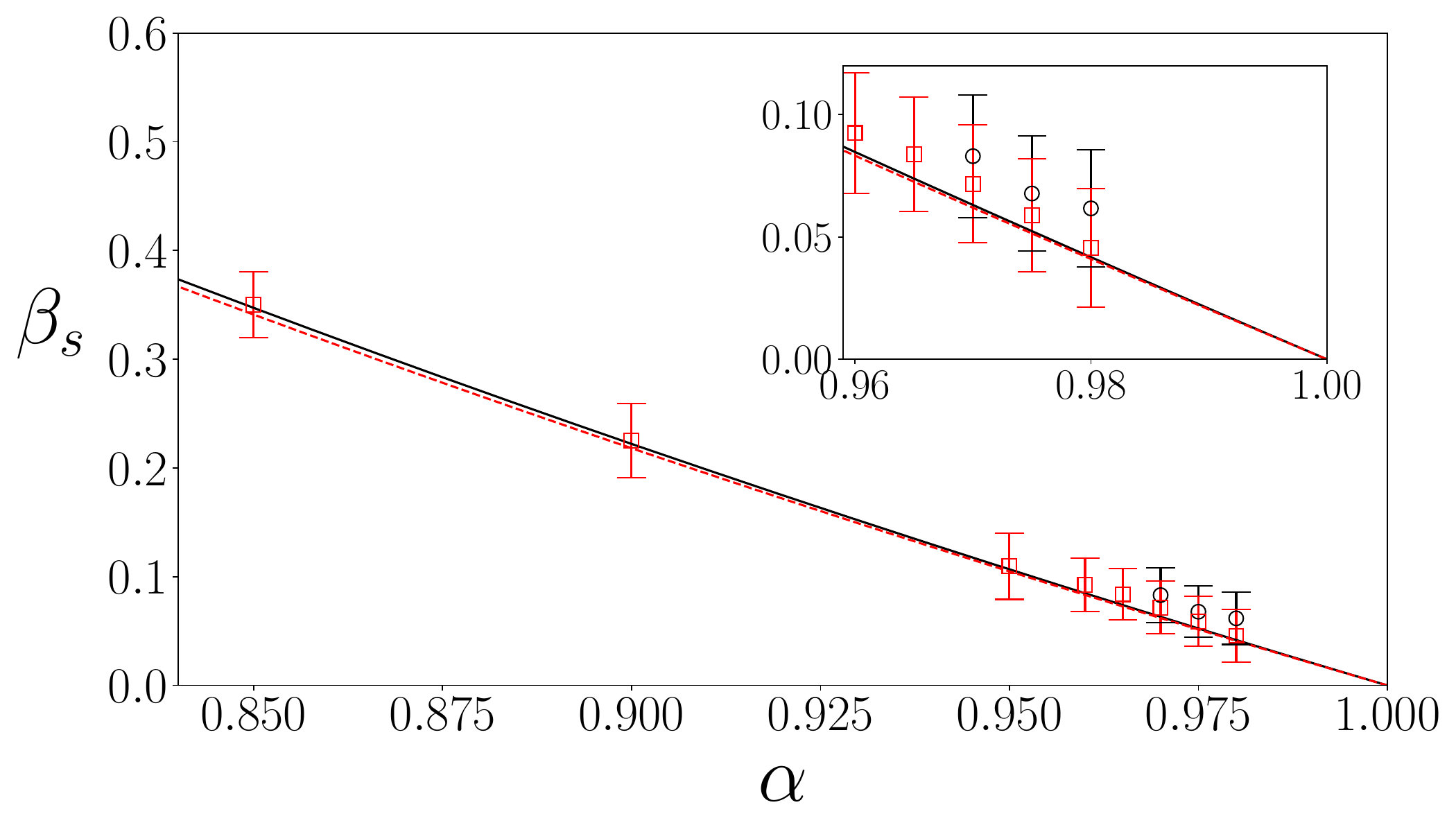}
 \caption{(Color online) $\beta_s$ of a hard disks system as a function
   of $\alpha$ for $H=10\sigma$ and 
$H=29\sigma$. The (black) circles are the simulation
results for $H=10\sigma$ 
and the (red) squares for $H=29\sigma$. The error bars are obtained from the average
over the $10$ realizations. The (black) solid line is the theoretical
prediction for $H=10\sigma$ and the (red) dashed line for $H=29\sigma$, both
given by Eq. (\ref{betaST}). The inset shows the quasi-elastic
region.   }
\label{betaSH1029}
\end{figure}

\begin{figure}[t!]
  \centering \includegraphics[scale=0.3]{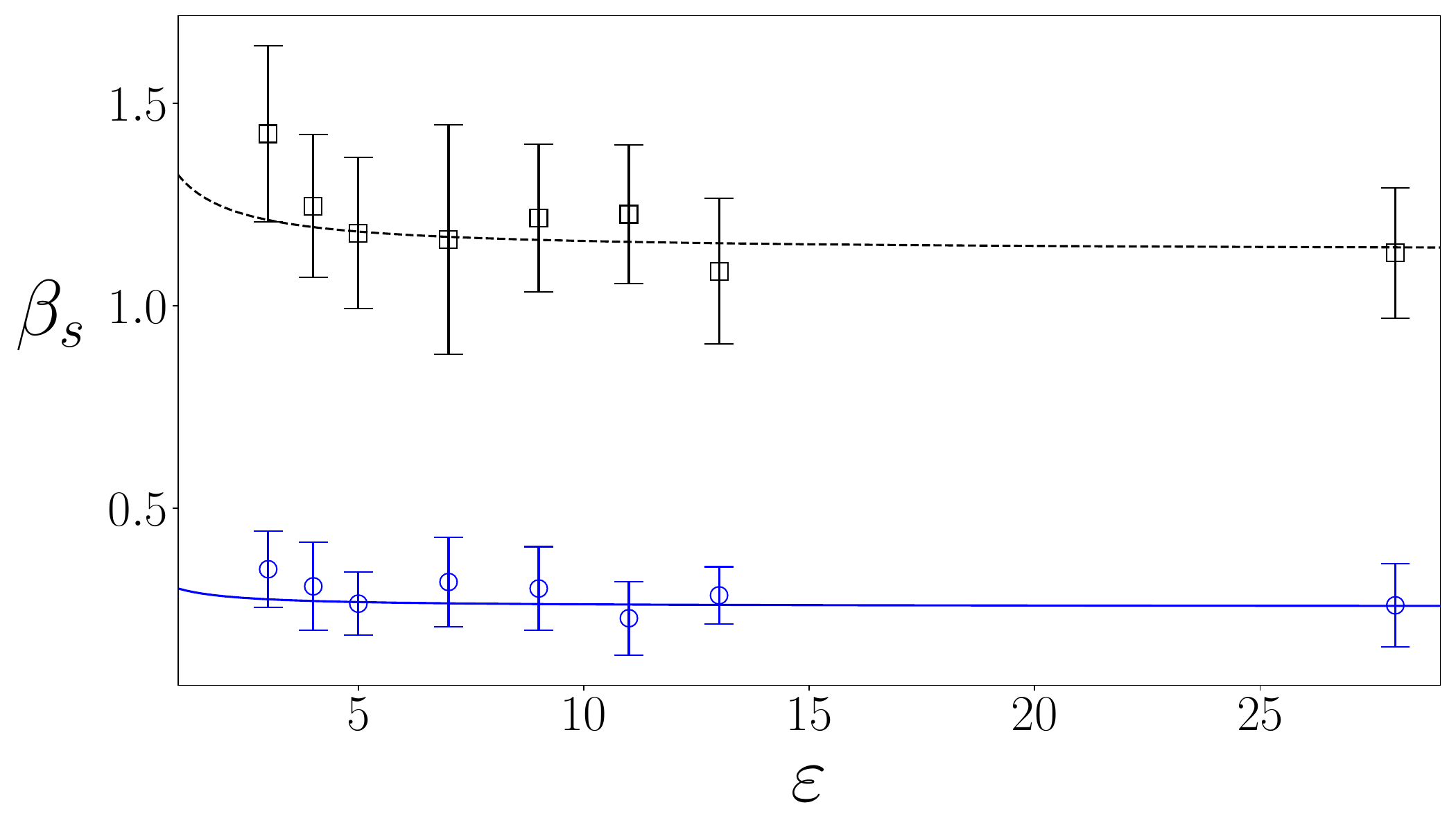}
 \caption{(Color online) $\beta_s$ of a hard sphere system as a function
   of the dimensionless height, $\varepsilon$, for $\alpha=0.6$ and 
$\alpha=0.9$. The (black) squares are the simulation
results for $\alpha=0.6$ 
and the (blue) circles for $\alpha=0.9$. The error bars are obtained from the average
over the $10$ realizations. The (black) point line is the theoretical
prediction for $\alpha=0.6$ and the (blue) solid line for $\alpha=0.9$, both
given by the numerical theoretical prediction coming
from Eq. (\ref{ecu28}). }
\label{betaVSHd3}
\end{figure}

In Fig. \ref{betaSH5}, the stationary  quotient of temperatures of a
system of hard spheres is plotted as a function of $\alpha$ for
$H=5\sigma$. The circles are the simulation results, the solid line
is the linear approximation of $\beta_s$ given by Eq. (\ref{betaST})
and the dashed line is the numerical theoretical prediction coming
from Eq. (\ref{ecu28}). The error bars are obtained from the average
over the $10$ realizations. The quasi-elastic region is enlarged in the
inset. It is seen that $\beta_s$ decays monotonically with the
inelasticity vanishing in the elastic limit, where equipartition
holds. A very good agreement is found between the
simulation results and the theoretical prediction coming from
Eq. (\ref{ecu28}) in the whole range of inelasticities. Stronger
inelasticities are not consider because the homogeneous state becomes
unstable.  The agreement
with Eq. (\ref{betaST}) is also good for mild inelasticities where
$\beta_s$ is still small, consistently with the linear
approximation. The same is plotted in Fig. \ref{betaSH29}, but for
$H=29\sigma$. As for $H=5\sigma$, a very good agreement is found
between the simulation results and the theoretical predictions. Note
that, for a given $\alpha$, the values of $\beta_s$ for $H=5\sigma$
and for $H=29\sigma$ are very close. This is in contrast with the
ultra-confined system where $\beta_s$ strongly depends on $H$
\cite{mgb19}. 
%-------------------------------
% ------- Imagen de beta estacionario 
For hard disks, a very similar behavior is obtained, but for a more
restricted range of inelasticities, because clusters of particles are
developed for larger values of $\alpha$. In Fig. \ref{betaSH1029}
(color online),
$\beta_s$ is plotted as a function of $\alpha$ for $H=10\sigma$ and
$H=29\sigma$. The (black) circles are the simulation results for $H=10\sigma$
and the (red) squares for $H=29\sigma$. The error bars are obtained from the average
over the $10$ realizations. The (black) solid line is the theoretical
prediction for $H=10\sigma$ and the (red) dashed line for $H=29\sigma$, both
given by Eq. (\ref{betaST}). The inset shows the quasi-elastic
region. In this case, it is not necessary to go
beyond the linear approximation as the range of inelasticities is
restricted to mild inelasticities. In fact, for $H=10\sigma$, clusters
are developed for $\alpha<0.975$. In Fig. \ref{betaVSHd3}, the
dependence of $\beta_s$ with the dimensionless height, $\varepsilon$,
is studied in a hard sphere system for two different values of the
inelasticity. The (black) squares are the simulation results for $\alpha=0.6$ 
and the (blue) circles for $\alpha=0.9$. The error bars are obtained
from the average over the $10$ realizations. The (black) point line is
the theoretical prediction for $\alpha=0.6$ and the (blue) solid line
for $\alpha=0.9$, both given by the numerical theoretical prediction coming
from Eq. (\ref{ecu28}). A good agreement between the theoretical
prediction and the simulation results is obtained. For $\varepsilon\sim
1$, it is appreciated
that the small deviation of $\beta_s$ with respect to the ``bulk''
contribution is captured by the theory.  Similar results are obtained
for hard disks. 

\begin{figure}[t]
\begin{center}
    \begin{subfigure}{0.49\textwidth}
\centering \includegraphics[scale=0.30]{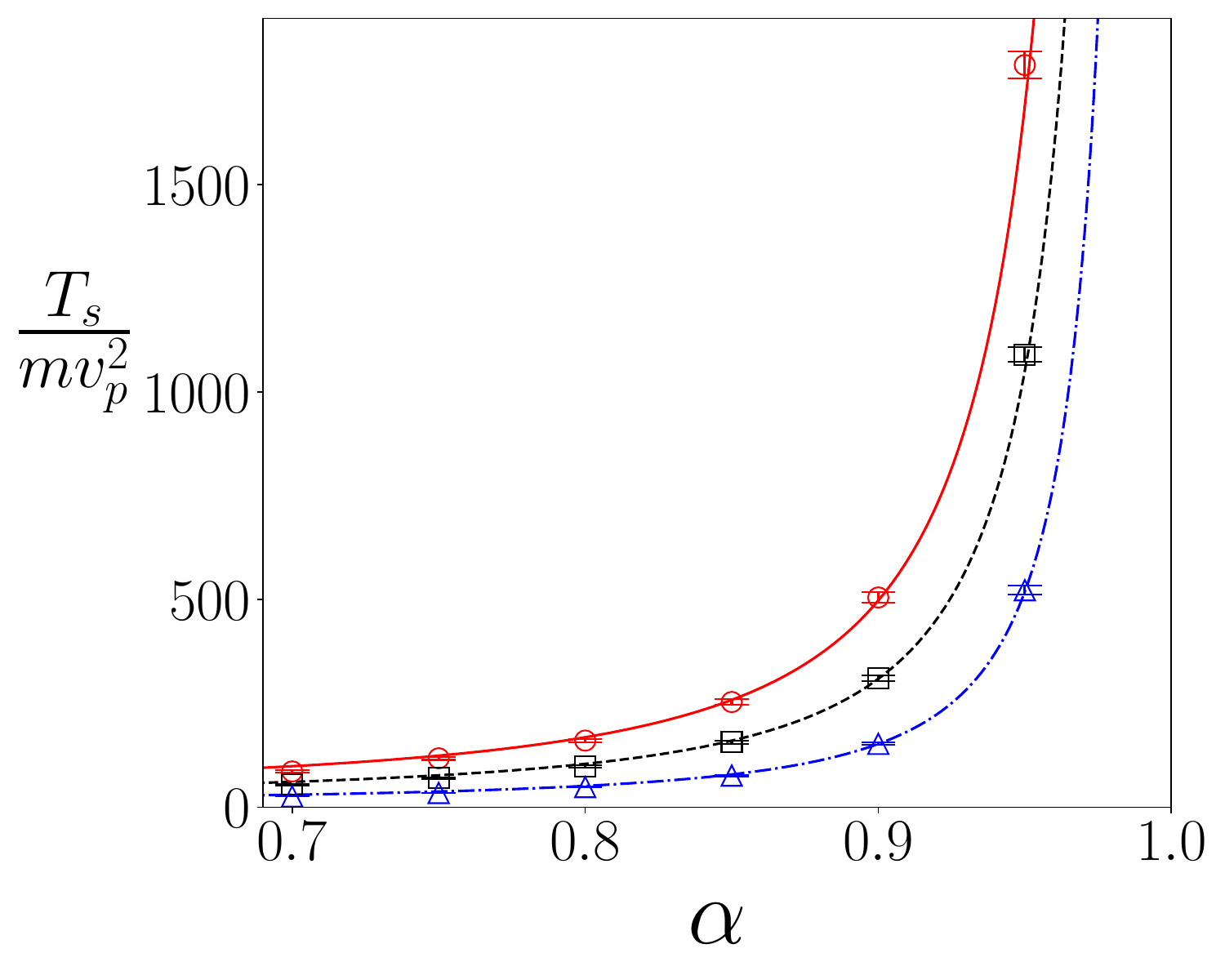}
    \caption{}
    \label{t_vs_alpha3d1}
\end{subfigure}
    \begin{subfigure}{0.49\textwidth}

    \centering \includegraphics[scale=0.30]{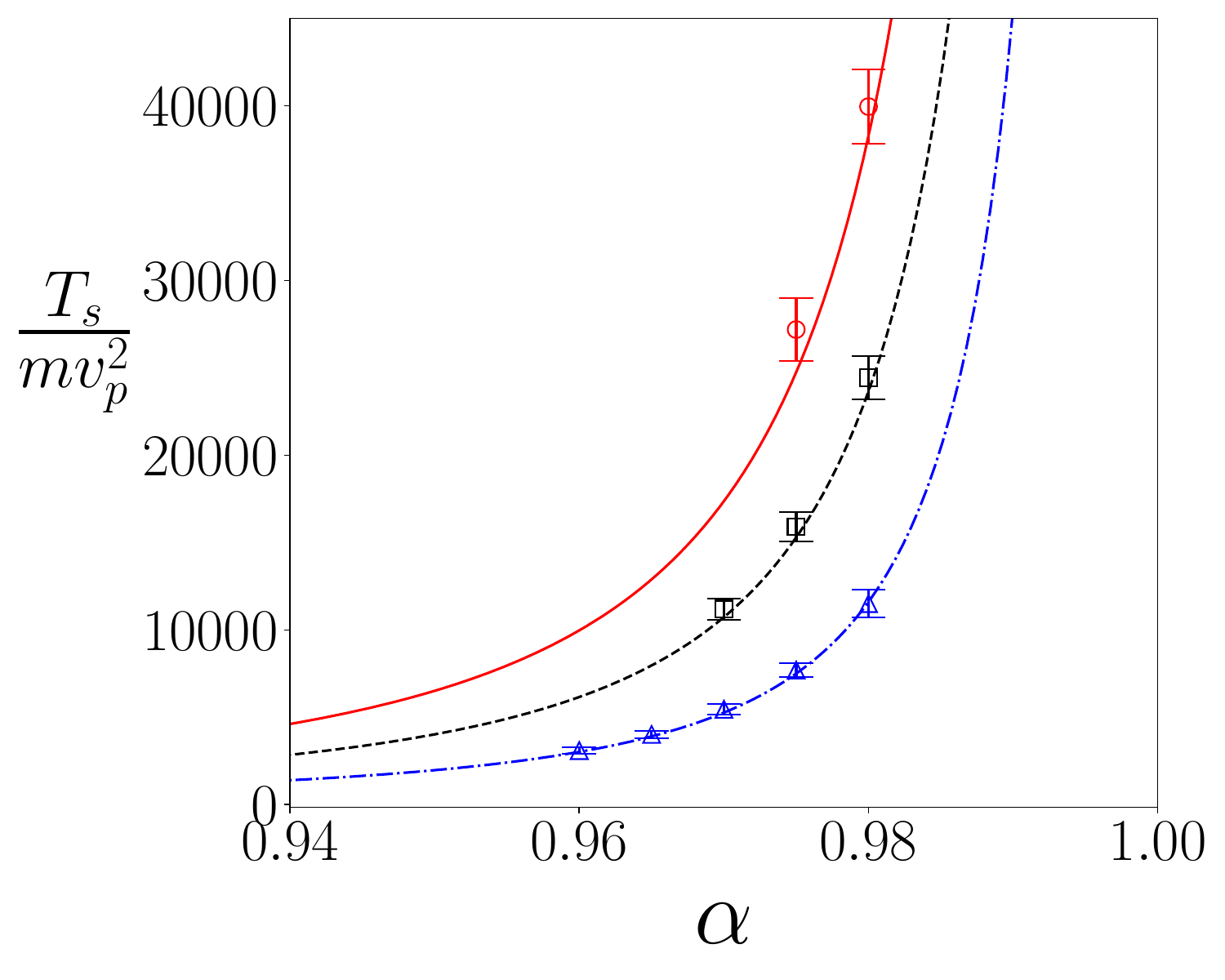}
    \caption{}
    \label{t_vs_alpha2d1}
\end{subfigure}

\end{center}

\caption{(Color online) Dimensionless horizontal stationary temperature,
$\frac{T_s}{mv_p^2}$, for 
hard spheres (a) and disks (b) as
a function of the inelasticity for different values of the height. The
(red) circles, (black) squares and (blue) triangles are the simulation results for
$H=8\sigma$, $H=10\sigma$ and $H=14\sigma$, respectively. The error
bars are obtained from the average over the $10$ realizations. The (red) solid
line, (black) dashed line and (blue) solid-dashed line are the corresponding
theoretical predictions given by Eq. (\ref{TsTheory}).  }
\label{Txy_vs_alpha}
\end{figure}

\begin{figure}[t!]
  \centering \includegraphics[scale=0.3]{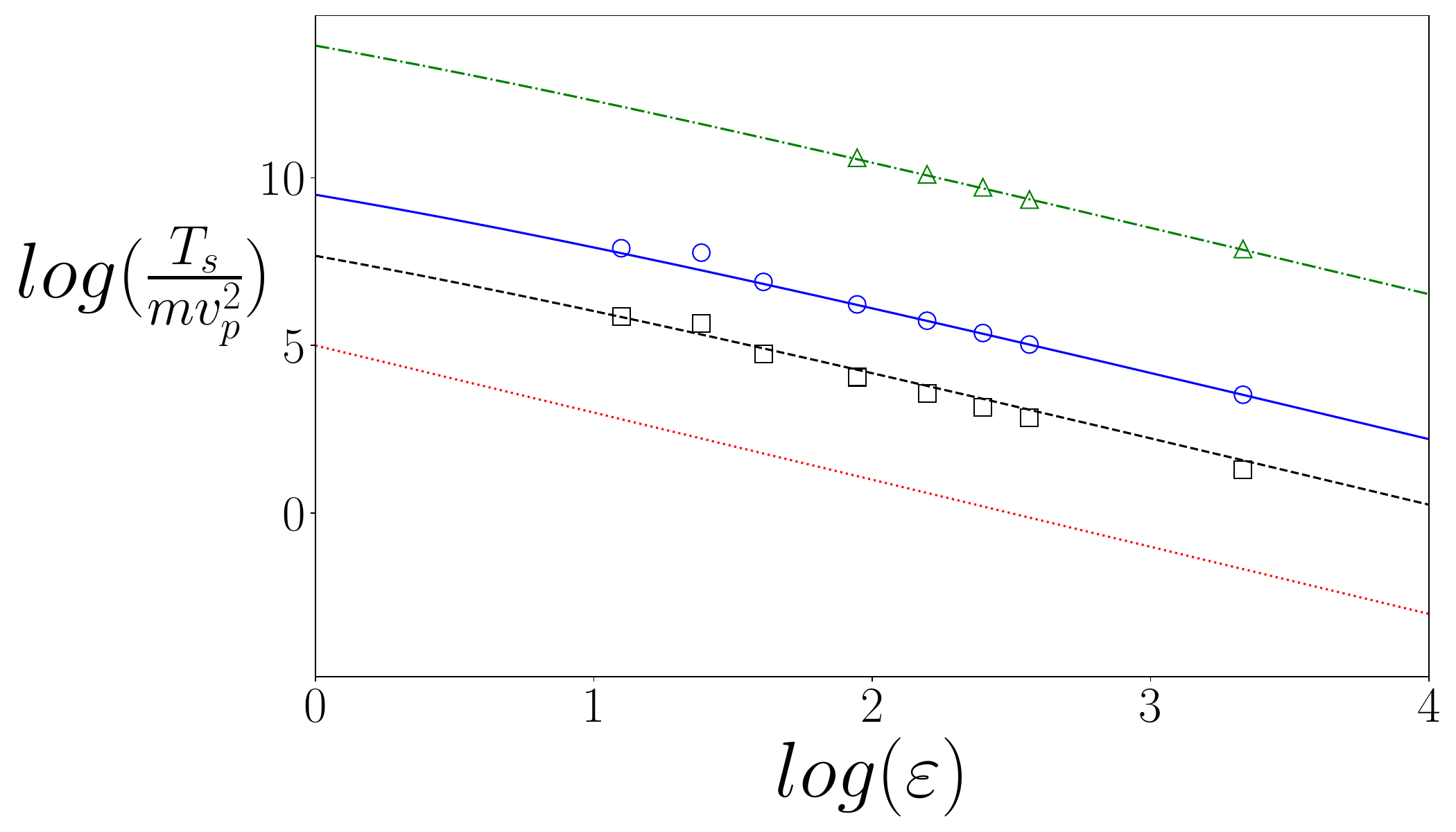}
 \caption{(Color online) Dimensionless
  horizontal stationary temperature, 
$\frac{T_s}{mv_p^2}$, as a function of $\varepsilon$ in logarithmic scale. The
(blue) circles and the (black) squares are hard spheres simulation results for
$\alpha=0.9$ and $\alpha=0.6$ respectively. The (green) triangles are
hard disks simulation results for $\alpha=0.98$. The 
(blue) solid line, (black) dashed line and (green) solid-dashed line are 
the corresponding theoretical prediction given by
Eq. (\ref{TsTheory}). The (red) point line is a straight line
with slope $-2$.}
\label{lnTxy_vs_h}
\end{figure}

The dimensionless horizontal stationary temperature,
$\frac{T_s}{mv_p^2}$, is plotted in Figs. \ref{t_vs_alpha3d1} and
\ref{t_vs_alpha2d1} (color online) for 
hard spheres and disks respectively as
a function of the inelasticity for different values of the height. The
(red) circles, (black) squares and (blue) triangles are the simulation results for
$H=8\sigma$, $H=10\sigma$ and $H=14\sigma$, respectively. The error
bars are obtained from the average over the $10$ realizations. The (red) solid
line, (black) dashed line and (blue) solid-dashed line are the corresponding
theoretical predictions given by Eq. (\ref{TsTheory}). The numerical
solution valid beyond the linear regime in $\beta_s$ is not plotted
because no difference between both curves can be appreciated. For a
given height, the
horizontal temperature increases with $\alpha$ because there is less dissipation of
energy in particle-particle collisions. For a given inelasticity, the
horizontal temperature decreases with the height as there are less particle-wall
collisions. The
agreement between the simulation results and the theoretical
prediction is excellent for the whole range of considered
parameters. In Fig. \ref{lnTxy_vs_h} (color online), the dimensionless
  horizontal stationary temperature, 
$\frac{T_s}{mv_p^2}$, is plotted as a function of $\varepsilon$ in
logarithmic scale. The
(blue) circles and the (black) squares are hard spheres simulation results for
$\alpha=0.9$ and $\alpha=0.6$ respectively. The (green) triangles are
hard disks simulation results for $\alpha=0.98$. The error bars are
not plotted because they can not be seen in the scale of the figure. The 
(blue) solid line, (black) dashed line and (green) solid-dashed line are 
the corresponding theoretical prediction given by
Eq. (\ref{TsTheory}). The (red) point line is a straight line
with slope $-2$ (showing the ``bulk'' prediction) that is plotted only
for
reference. It is seen that the
agreement between the theory and the simulations is very good and that
the corrections with respect to the bulk prediction are even smaller than for
$\beta_s$ in the whole range of heights. No more values of the
inelasticity are considered for hard disks because, for smaller values
of the inelasticity, the system is unstable for most of the values of
$\varepsilon$. 
%-------------------------------
% ------- Imagen de temperaturas estacionarias 

\begin{figure}[t]
\begin{center}
    \begin{subfigure}{0.5\textwidth}

  \centering \includegraphics[scale=0.28]{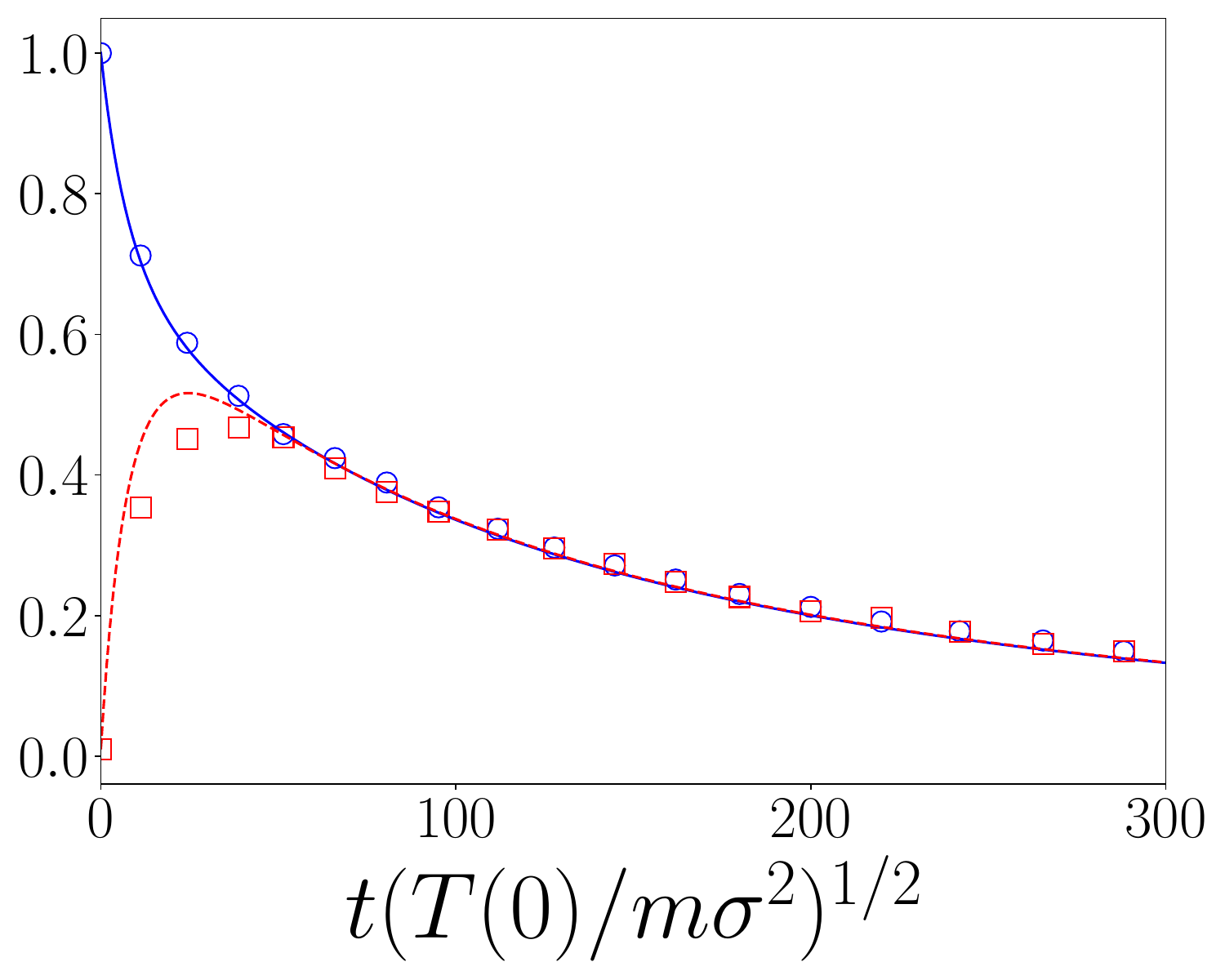}
    \caption{}
    \label{temp_vs_t3d}
\end{subfigure}
\hfill
\begin{subfigure}{0.49\textwidth}
 
\centering \includegraphics[scale=0.28]{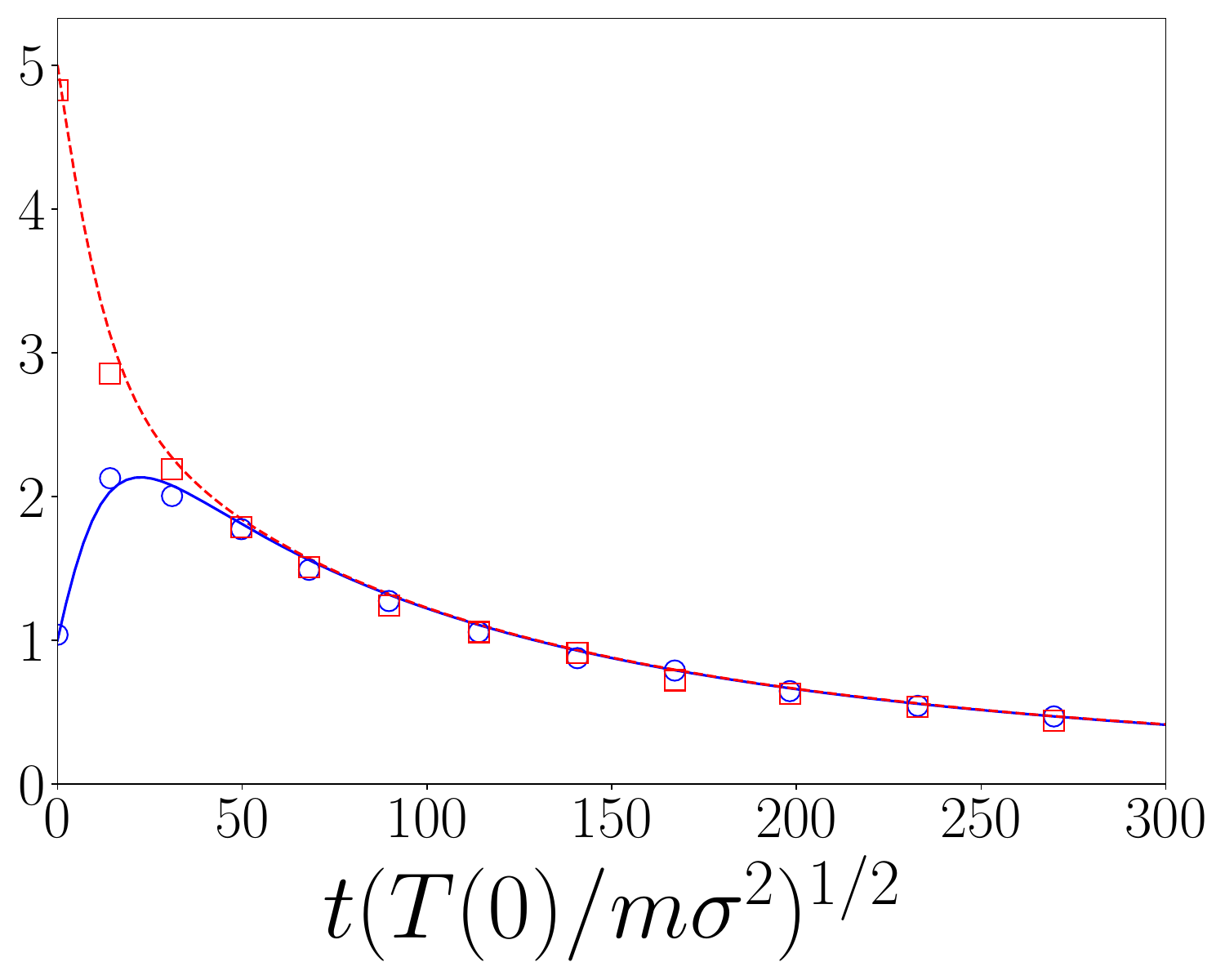}
    \caption{}
    \label{temp_vs_t2d}
\end{subfigure}
\end{center}

\caption{(Color online) Time 
evolution of the temperatures for a system of hard spheres (a) and
disks (b), as a function of the dimensionless time
$\left[\frac{T(0)}{m\sigma^2}\right]^{1/2}t$. In both cases, the values of the
parameters are $H=29\sigma$ and $\alpha=0.9$. The (blue) circles and the
(red) squares are the simulation results for $T$ and $T_z$,
respectively. The (blue) solid line and (red) dashed lines are the
corresponding theoretical prediction  
obtained by solving numerically Eqs. (\ref{ecu31}) and
(\ref{ecu32}). }
\label{T_vs_t}
\end{figure}

Finally, we have also studied the time evolution of the  horizontal and
vertical temperatures in order to test if Eqs. (\ref{ecu31}) and (\ref{ecu32})
describe correctly the dynamics. In Fig. \ref{temp_vs_t3d} and
\ref{temp_vs_t2d} (color online), the time 
evolution of the temperatures for a system of hard spheres and disks,
respectively, is plotted as a function of
the dimensionless time
$\left[\frac{T(0)}{m\sigma^2}\right]^{1/2}t$. In both cases, the values of the
parameters are $H=29\sigma$ and $\alpha=0.9$ and the initial condition
verifies $T_z(0)=0.01T(0)$ and $T_z(0)=5T(0)$ for hard spheres and
disks, respectively. The (blue) circles and the
(red) squares are the simulation results for $T$ and $T_z$,
respectively. The (blue) solid and (red) dashed lines are the corresponding
theoretical prediction  
obtained by solving numerically Eqs. (\ref{ecu31}) and
(\ref{ecu32}). The agreement between the theoretical prediction and
the simulation results is excellent for the whole time window. Similar
results are obtained for different initial conditions and/or different
values of the parameters if the system remains spatially
homogeneous. Hence, we can conclude that Eqs. (\ref{ecu31}) and
(\ref{ecu32}) describe accurately the dynamics of the partial
temperatures for spatially homogeneous states. 

%-------------------------------
% ------- Imagen de dinámica 

\section{Discussion and Conclusions}
\label{SecV}

In this work, we have formulated a kinetic equation for a dilute granular
system composed of inelastic hard spheres or disks that are confined
between two vertically vibrating walls separated a distance larger
than twice the diameter of the particles. The equation for the
one-particle distribution function has the typical free-streaming part
and the collisional contribution that takes into account the
particle-particle and the particle-wall collisions. The
particle-particle collisional term takes into account the effects of
the confinement by restricting the orientation of collisions to such
in which the two particles involved are inside the system. The spatial
domain of the system is naturally divided into the ``bulk'' part (in
which all the orientations of the collisions are allowed) plus the
``boundary'' part closed to the walls (where the orientation of the
collisions are restricted). Although the kinetic equation makes sense
for arbitrary height, $H$, we have restricted ourselves to small
  enough heights so that the
assumption that the distribution function is $z$-independent is
expected to be valid. In this case, a closed evolution equation for
the marginal distribution, $f(\bm{r}_{\parallel}, \bm{v}, t)
   \equiv \frac{1}{(H-\sigma)} \int_{\sigma/2}^{H-\sigma/2} \dd{z} ~
   f(\bm{r}, \bm{v}, t)$, is obtained and, remarkably, the
   particle-particle collisions contribution splits into two terms:
   one corresponding to an ultra-confined system of height $2\sigma$, plus
   other corresponding to a bulk system (there are not restrictions on
   the orientation of the collisions) of height $H-2\sigma$. 

Considering that there are not gradients in the horizontal direction,
the kinetic equation is solved by assuming that the distribution
function is a gaussian with two temperatures (the horizontal and
vertical temperatures). Closed evolution equations for the partial
temperatures are obtained, that are explicitly written in the linear
in $\beta\equiv\frac{T_z}{T}-1$ approximation, Eqs. (\ref{ecu31}) and 
(\ref{ecu32}). The structure of the equations is transparent: energy
is injected in the vertical direction due to the walls and it is
transferred to the horizontal direction through collisions. Energy is
dissipated in particle-particle collisions that is reflected in
both equations. The equations admit a stationary state in which the
energy lost in collisions is compensated by the energy injected by the
particle-wall collisions. Moreover, we have shown that this solution
is linearly stable. The temperature quotient in the stationary state,
$\beta_s$, is always positive (indicating that $T_{z, s}>T_s$) and
decays monotonically with $\alpha$, vanishing in the elastic limit
(consistently with 
equipartition). On the other hand, the horizontal temperature,
$T_s$, decays with the height (consistently with the fact that there
are less particle-wall collisions with respect to particle-particle
collisions) and increases with $\alpha$, as expected. While $\beta_s$
is density independent, $T_s\sim (n\sigma^d)^{-2}$. This behavior can be
intuitively understood from the fact that energy injection comes
from particle-wall collisions and goes with $n\sigma^d$, while
dissipation comes from particle-particle collisions that goes with
$(n\sigma^d)^2$. The theoretical predictions that comes from Eqs. (\ref{ecu31}) and 
(\ref{ecu32}) agree very well with MD simulation 
results for a wide range of the parameters. We have shown that for
mild inelasticities, let us say $0.8<\alpha<1$, the agreement is
excellent, both for the stationary values and for the dynamics. For
stronger inelasticities, the agreement is also very good when the
equations beyond the linear approximation in $\beta$ are numerically
solved. All these results give a strong support to the kinetic theory
developed in the paper.  

\begin{figure}[t!]
  \centering \includegraphics[scale=0.18]{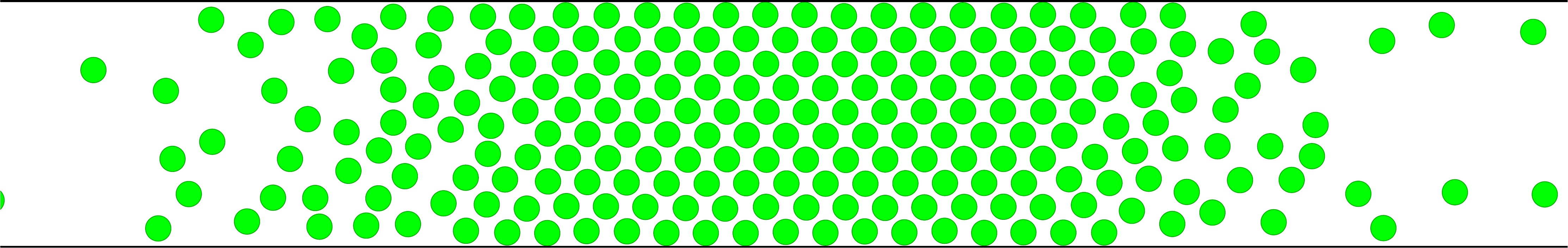}
 \caption{Snapshot of a hard disks system once
the stationary state has been reached. Only a portion of
the system close to the region where
the inhomogeneities are developed is plotted. The values of the parameters
are the ones for hard disks with $H=8\sigma$ and $\alpha=0.85$. }
\label{figSnaphotCluster}
\end{figure}

The obtained density dependence of the horizontal stationary
temperature is relevant, as the density dependence of the stationary
pressure is, then, $p_s\sim(n\sigma^d)^{-1}$. We have already
mentioned that the fact that $\frac{dp_s}{dn}<0$ is related with the
presence of an instability. In fact, in the simulations, clusters were
developed for some values of the parameters that, of course, were
disregarded in the previous section in order to compare with the
theoretical predictions. In 
Fig. \ref{figSnaphotCluster}, a snapshot of a hard disks system once
the stationary state has been reached is shown. Only a portion of
the system close to the region where
the inhomogeneities are developed is plotted. The cluster of particles 
can be clearly seen. The values of the parameters
are the ones for hard disks with $H=8\sigma$ and $\alpha=0.85$.  
Let us remark that this instability is compatible with the fact that
the stationary solution of  Eqs. (\ref{ecu31}) and 
(\ref{ecu32}) is linearly stable, as in the stability analysis the spatial
homogeneity was taken for granted. In order to analyze the nature of
the present instability, a similar analysis to the one performed in
\cite{mgb19b} could be done, i.e., to perform a linear stability
analysis of the homogeneous state by assuming a hydrodynamic 
description in the plane. Nevertheless, in Ref. \cite{mgb19b} it was
clear that the two-dimensional transport coefficients should be taken
(at least for $\varepsilon\ll 1$), whether in our case, it is not so 
clear. In fact, here it becomes evident that, as long as the kinetic
equation for ultra-confined systems \cite{bmg16} is
the natural tool to tackle the transition from a three-dimensional
system to a two-dimensional one, the kinetic equation of the
paper, Eq. (\ref{ecu5}) or Eq. (\ref{ecu15}), let us study the
transition from a confined system to a ``pure'' three-dimensional system
where the ``boundary'' terms of the kinetic equation are not relevant. 

Finally, let us mention that Eqs. (\ref{ecu31}) and (\ref{ecu32})
describe the time evolution of the partial temperatures 
in a very general framework. The only essential ingredient is that the 
ansatz given by Eq. (\ref{ecu22}) be approximately valid. Hence, the
equations can be the starting point of further 
studies such as the study of the Kovacs \cite{pt14, bgmb14} or Mpemba
\cite{lvps17} effects. The difference with the models of the above mentioned
references is that, in our context, the variables involved in the process
can be controlled and the effects can be experimentally
relevant. In \cite{bpr22}, the Mpemba effect is studied in a more
  similar model in which the energy is injected anisotropically,
  although it is hard to perform 
a direct comparison between the results for both models as the
energy injection 
mechanisms are different (it is stochastic in the case of \cite{bpr22}, 
while it is deterministic in ours). Beyond these transient effects,
from Eqs. (\ref{ecu31}) and 
(\ref{ecu32}), it would also be possible to analyze if, before the stationary
state is reached, a universal state (in the sense that it is
independent of the initial condition) is reached in the long-time
limit as it happens in other granular systems \cite{gmt12,
  bmgb14}. This point is specially relevant if a hydrodynamic in the
plane description is possible. Work in these lines is in progress.

\section*{Acknowledgments}

We thank J. J. Brey for fruitful discussions and for a careful reading
of the manuscript. This research was supported by grant
ProyExcel-00505 funded by Junta de 
Andaluc\'ia, and grant PID2021-126348N funded by
MCIN/AEI/10.13039/501100011033 and "ERDF A way of making Europe".

%****************************************************************************************************************************
\appendix

  \section{Properties of the particle-wall collision term}
  \label{Apendice_A1}
In this appendix we will proof the relations \eqref{ecu22}, and \eqref{ecu23}. To do that, we start using the expression of the operator $L_{\uparrow}$ defined in \eqref{ecu10} in order to develop explicitly the left hand side of \eqref{ecu22}
\begin{align}\label{ecua1}
\begin{aligned}
     \int \dd \bm{v} \psi(\bm{v})L_{\uparrow} f(\bm{v},t) = & \int \dd \bm{v} \psi(\bm{v})   \left[
  \Theta(-v_{z}-2v_p) |2v_p+v_z|b_{\uparrow}- \Theta(v_{z}) v_{z} \right] f( \bm{v}, t),  \\
  = &   \int \dd \bm{v} \psi(\bm{v})   \Theta(b_{\uparrow} v_z ) |b_{\uparrow}v_z| f(  b_{\uparrow} \bm{v}, t)  - \int \dd \bm{v}  \psi(\bm{v}) \Theta(v_{z}) v_{z}  f( \bm{v}, t)  ,
  \end{aligned}
\end{align}
when we use the definition of $b_{\uparrow}$ operator. Making a change of variable ${\bm u } \equiv  b_{\uparrow}{\bm v}$ in the first integral of the right hand site, and taking into account that the jacobian is the unity, we have 
\begin{align}\label{ecua2}
\begin{aligned}
     \int \dd \bm{v} \psi(\bm{v})L_{\uparrow} f(\bm{v},t)   = &   \int \dd \bm{u} \psi(b_{\uparrow}\bm{u})   \Theta( u_z ) |u_z| f(  \bm{u}, t)  - \int \dd \bm{v}  \psi(\bm{v}) \Theta(v_{z}) v_{z}  f( \bm{v}, t)  , \\
       = &   \int \dd \bm{v}  f(  \bm{v}, t)    \abs{v_z}  \Theta( v_z )  (b_{\uparrow}-1)    \psi(\bm{v}) .
  \end{aligned}
\end{align}

Note that the property $b_{\uparrow}^{-1} = b_{\uparrow} $ is used in the last equality. Using the same steps, the expression \eqref{ecu23} is obtained. 

\section{Evaluation of the collision integrals}
\label{Apendice_B1}

The objective in this Appendix is to calculate the particle-particle
collision contribution to the evolution equations of the partial
temperatures. Taking into account Eq. (\ref{ecu25}) and 
\begin{align}\label{ecub1}
\begin{aligned}
  (b_{\widehat{\bm\sigma}}-1)(v_z^2+v_{1z}^2)=& v_{z}'^2+ v_{1z}'^2-v_z^2-v_{1z}^2 ,\\
  =& \frac{(\alpha +1)^2}{2}(\bm{g}\cdot \widehat{\bm\sigma})^2
  \widehat{\sigma}_z^2- (\alpha +1)(\bm{g}\cdot\widehat{\bm\sigma} )
  \widehat{\sigma}_z 
  g_z, 
\end{aligned}
\end{align}
we have
 \begin{align}   \label{ecub2}
 \begin{aligned}
\frac{1}{H-\sigma}\int_{\sigma/2}^{H-\sigma/2}dz\int \dd \bm{v} v_z^2
\Jz=
\frac{\sigma^{d-1}(\alpha
  +1)}{2(H-\sigma)}\int_{\sigma/2}^{H-\sigma/2} \dd z \int \dd 
\bm{v}_1 \int \dd \bm{v} f(\bm{v}_1,t) f(\bm{v},t) \\
\int_{\Omega_d(H, z)} \dd \widehat{\bm \sigma}\abs{\bm{g} \cdot
  \widehat{ \bm \sigma}}\Theta(\bm{g}\cdot \widehat{ \bm \sigma} ) 
\left[ 
\frac{(\alpha +1)}{2}(\bm{g}\cdot \widehat{\bm \sigma})^2 \widehat{\sigma}_z^2-
(\bm{g}\cdot \widehat{\bm \sigma}) \widehat{\sigma}_z g_z  \right]\\ 
= \frac{\sigma^{d-1}(\alpha
  +1)}{2(H-\sigma)}\int_{\sigma/2}^{H-\sigma/2} \dd z\int \dd \bm{v}_1
\int \dd 
\bm{v} f(\bm{v}_1,t) f(\bm{v},t) \int_{\Omega_d(H, z)} \dd
\widehat{\bm \sigma}\abs{\bm{g} \cdot \widehat{ \bm
    \sigma}}\Theta(-\bm{g} \cdot \widehat{ \bm \sigma} )\\ 
\left[
\frac{(\alpha +1)}{2}(\bm{g}\cdot \widehat{\bm \sigma})^2
\widehat{\sigma}_z^2-(\bm{g}\cdot \widehat{\bm \sigma})
\widehat{\sigma}_z g_z  \right], 
\end{aligned}
 \end{align}
where we have changed the labels $\bm{v}\leftrightarrow \bm{v}_1$ in the last step.
In this way, we can integrate over all values of  $\bm{v}_1$ and
$\bm{v}$ (without the restriction coming from $\Theta$) and we have
\begin{align}\label{ecub3}\begin{aligned}
 \frac{1}{H-\sigma}\int_{\sigma/2}^{H-\sigma/2}dz\int \dd \bm{v} v_z^2
 \Jz
=\frac{\sigma^{d-1}(\alpha +1)}{4(H-\sigma)}
\int_{\sigma/2}^{H-\sigma/2} \dd z\left [\frac{(\alpha +1)}{2} 
  H_1(z)-H_2(z)\right ], 
\end{aligned}
\end{align}
 with 
\begin{eqnarray}
H_1(z) & \equiv &\int \dd \bm{v}_1 \int \dd \bm{v} f(\bm{v}_1,t)
f(\bm{v},t) \int_{\Omega_d(H, z)} \dd \widehat{\bm \sigma}\abs{\bm{g}
  \cdot \widehat{ \bm \sigma}}^3 \widehat{\sigma}_z^2,\\ 
H_2(z) & \equiv &\int \dd \bm{v}_1 \int \dd \bm{v} f(\bm{v}_1,t)
f(\bm{v},t) \int_{\Omega_d(H, z)} \dd \widehat{\bm \sigma}\abs{\bm{g}
  \cdot \widehat{ \bm \sigma}} (\bm{g}\cdot \widehat{\bm \sigma})
\widehat{\sigma}_z g_z . 
\end{eqnarray}

Let us first calculate the contribution coming from $H_1(z)$. Taking
into account the shape of the distribution function given by
Eq. \eqref{ecu22}, we make the change of variables 
 \begin{align}
\bm{C} &= \frac{1}{2}(\bm{k} + \bm{k}_1),  \label{ecub5}\\
\bm{c} &= \bm{k}_1 - \bm{k}, \label{ecub6}
 \end{align}
where 
\begin{align}
\bm{k} &=\frac{v_x}{w}\widehat{ \bm
         e}_x+\frac{v_y}{w}\widehat{ \bm
         e}_y+\frac{v_{z}}{w_z} \widehat{ \bm e}_z,  \\ 
\bm{k}_1 &= \frac{v_{1x}}{w}\widehat{ \bm
         e}_x+\frac{v_{1y}}{w}\widehat{ \bm
         e}_y+\frac{v_{1z}}{w_z}\widehat{ \bm e}_z, 
 \end{align} 
for $d=3$ and 
\begin{align}
\bm{k} &=\frac{v_y}{w}\widehat{ \bm
         e}_y+\frac{v_{z}}{w_z} \widehat{ \bm e}_z,  \\ 
\bm{k}_1 &= \frac{v_{1y}}{w}\widehat{ \bm
         e}_y+\frac{v_{1z}}{w_z}\widehat{ \bm e}_z, 
 \end{align} 
for $d=2$. 
Note that the jacobian of transformation is $\abs{\mathbb{J}}=
w^{2(d-1)} w_z^2$.  Taking into account Eq \eqref{ecu22}, $f(\bm{v}_1,t)
f(\bm{v},t)$ is expressed in the new variables
 \begin{align}\label{ecub7}
 f(\bm{v}_1,t) f(\bm{v},t) = \Bigg[  \frac{n}{\pi^{\frac{d}{2}}
   w^{d-1}(t) w_z(t)} \Bigg]^2\exp(-2 C^2)\exp(-\frac{1}{2}c^2). 
\end{align}  
In these variables, $H_1(z)$ is expressed as  
 \begin{align} \label{ecub9}
\begin{split}
H_1(z)=\Bigg(\frac{n}{\pi^{\frac{d-1}{2}}}\Bigg)^2 \int \dd
\bm{C}\exp(-2 C^2) \int \dd \bm{c} \exp(-\frac{1}{2}c^2)
\int_{\Omega_d(H, z)} \dd \widehat{\bm \sigma}\abs{\bm{c} \cdot \widehat{ \bm a}}^3\\
  \Big[(w \widehat{\sigma}_\parallel)^2+ (w_z \widehat{\sigma}_z)^2  \Big]^{3/2}
 \sigma_z^2, 
\end{split}
 \end{align}
where
$\widehat{\sigma}_\parallel\equiv|\widehat{\bm{\sigma}}-\widehat{\sigma}_z\widehat{e}_z|$
and $\widehat{ \bm a}$ is a unitary vector in the direction of
$w\widehat{\sigma}_x\widehat{\bm{e}}_x+w\widehat{\sigma}_y\widehat{\bm{e}}_y
+w_z\widehat{\sigma}_z\widehat{\bm{e}}_z$ or 
$w\widehat{\sigma}_y\widehat{\bm{e}}_y
+w_z\widehat{\sigma}_z\widehat{\bm{e}}_z$ for $d=3$ and $d=2$
respectively. The  integral in $\bm{C}$ is trivial
 \begin{align} \label{ecub10}
 \int \dd \bm{C}\exp(-2 C^2) = \left(\frac{ \pi}{2}\right)^{d/2}, 
 \end{align} 
and the integral in $\bm{c}$ is
 \begin{align} \label{ecub11}
 \int  \dd \bm{c} \exp(-\frac{1}{2}c^2) \abs{\bm{c} \cdot \widehat{
   \bm a}}^3 = 4(2\pi)^{\frac{d-1}{2}}, 
 \end{align}
 where we have used that the integral does
 not depend on $\widehat{ \bm a}$. 
In this way Eq. \eqref{ecub9} reduces to
\begin{align} \label{ecub14} \begin{aligned}
H_1(z)=&\frac{2 \sqrt{2} n^2w^3}{ \sqrt{ \pi}} \int_{\Omega_d(H, z)} \dd
\widehat{\bm \sigma}\left[  \widehat{\sigma}_\parallel^2+  \widehat{\sigma}_z^2+
  \left(\frac{w_z^2}{w^2}-1\right) \widehat{\sigma}_z^2  \right]^{3/2}
\sigma_z^2, \\ 
=& \frac{2 \sqrt{2} n^2w^3 }{ \sqrt{\pi }} \int_{\Omega_d(H, z)} \dd
\widehat{\bm \sigma}\left( 1+   \beta \widehat{\sigma}_z^2  \right)^{3/2}
\widehat{\sigma}_z^2, 
\end{aligned}
\end{align}
where we have defined $\beta \equiv (w_z^2/w^2-1) \equiv (T_z/T-1)$
and the equality $||\widehat{\bm \sigma}||=1$ has been
used. Considering spherical coordinates in $d$-dimensions and taking
into account the explicit parametrization of the domain
$\Omega_d(H,z)$, we have 
\begin{align} \label{ecub.18}
    \begin{aligned}
\frac{1}{(H-\sigma)} \int_{\sigma/2}^{H-\sigma/ 2} \dd z H_1(z)=
\frac{2  \sqrt{2} n^2w^3 \Omega_{T,d-1} }{   \sqrt{\pi} 
  \varepsilon \sigma} \left[ I_1 + \sigma(\varepsilon-1) I_2\right],
\end{aligned}
\end{align}
where we have defined
\begin{align}
    I_1 \equiv& \int_{\sigma/2}^{3\sigma/2} \dd z
                \int_{\pi/2-b_2(z)}^{\pi/2+ b_1(z)} \dd \theta \left(
                1+   \beta  \cos^2\theta  \right)^{3/2}
                \sin^{d-2}\theta\cos^2\theta,  \label{ecu_I1} \\
    I_2 \equiv & \int_{0}^{\pi} \dd \theta \left( 1+   \beta
                 \cos^2\theta  \right)^{3/2} \sin^{d-2}\theta 
\cos^2\theta.  \label{ecu_I2}
\end{align} 
 The first integral defined in Eq. \eqref{ecu_I1} can be simplified by
 changing variables from $\theta$ to $z_1$ defined through 
\begin{align}
\begin{aligned}\label{ecub.20}
    \cos\theta = \frac{z_1-z}{\sigma}.
\end{aligned}
\end{align}
It can be expressed as 
\begin{align} \label{ecub.21}
    \begin{aligned}
I_1 = & \frac{1}{\sigma} \int_{\sigma/2}^{3\sigma/2} \dd z
\int_{\sigma/2}^{3\sigma/2} \dd z_1 \left[ 1+   \beta
  \left(\frac{z-z_1}{\sigma}\right)^2   \right]^{3/2} \left[ 1-
  \left(\frac{z-z_1}{\sigma}\right)^2 \right]^{(d-3)/2} 
\left(\frac{z-z_1}{\sigma}\right)^2,
\end{aligned}
\end{align}
or, in terms of the dimensionless variables, 
\begin{align}
\tilde{z}_1=\frac{z_1-\sigma / 2}{\sigma}, \quad
  \tilde{z}_2=\frac{z-\sigma / 2}{\sigma},  
\end{align}
as
\begin{align}
  I_1 =   2 \sigma \int_0^{1} \mathrm{d} \tilde{z}_1 \int_0^{\tilde{z}_1} \mathrm{~d} \tilde{z}_2\left( 1+   \beta  \tilde{z}_{12}^2   \right)^{3/2} \left( 1- \tilde{z}_{12}^2 \right)^{(d-3)/2}
\tilde{z}_{12}^2,
\end{align}
where we have introduced $\tilde{z}_{12} \equiv
\tilde{z}_1-\tilde{z}_2$ and we have used that the integrand is
invariant under the change $\tilde{z}_1$ by $\tilde{z}_2$.  Finally,
changing variables to 
\begin{align}
    y=\tilde{z}_1-\tilde{z}_2, \quad Y=\frac{1}{2}\left(\tilde{z}_1+\tilde{z}_2\right),
\end{align}
it is obtained
\begin{align}
    \begin{aligned}\label{ecub.first_integral}
I_1 = 2 \sigma \int_0^1 \mathrm{d} y \int_{y / 2}^{1-y / 2}
\mathrm{~d} Y  \;y^2 \left(1-y^2\right)^{(d-3)/2}   \left(1+\beta
  y^2\right)^{3 / 2} =&2 \sigma \int_0^1 \mathrm{d} y
\;y^2(1-y)\left(1-y^2\right)^{(d-3)/2}\\ 
& \left(1+\beta y^2\right)^{3 / 2}.
\end{aligned}
\end{align}
 The integral $I_2$ can be expressed in a
 similar way that $I_1$. In effect, by introducing the new variable, 
 \begin{align}
 \begin{aligned}
     x = \cos \theta, 
 \end{aligned}
     \end{align}
$I_2$ can be expressed as 
 \begin{align} \label{ecub.second_integral}
  I_2=   \begin{aligned} 2  \int_{0}^{1} \dd x  \; x^2   (1-x^2)^{(d-3)/2}
\left( 1+   \beta  x^2  \right)^{3/2}.
\end{aligned}
\end{align}
By introducing Eqs. \eqref{ecub.first_integral} and
\eqref{ecub.second_integral} into Eq.  \eqref{ecub.18}, it is obtained 
\begin{align} \label{ecu_H1}
    \begin{aligned}
\frac{1}{(H-\sigma)} \int_{\sigma/2}^{H-\sigma/ 2} \dd z H_1(z)=&
\frac{4  \sqrt{2} n^2w^3 \Omega_{T,d-1} }{   \sqrt{\pi} \varepsilon }  \int_0^1
\dd y \; y^2 (1-y^2)^{(d-3)/2} (1+\beta y ^2)^{3/2}
[(\varepsilon-1)+(1-y)]. 
\end{aligned}
\end{align}
 
To get a similar expression for 
$\frac{1}{(H-\sigma)} \int_{\sigma/2}^{H-\sigma/ 2} \dd z H_2(z)$, we
follow similar steps. By 
changing variables to the ones introduced in equations \eqref{ecub5}
and \eqref{ecub6}, we have 
\begin{align} \label{ecub15}
\begin{split}
H_2(z) =  \left(\frac{n}{\pi^{
\frac{d}{2}}}\right)^2 \int \dd \bm{C}\exp(-2 C^2) \int 
\dd \bm{c} \exp(-\frac{1}{2}c^2)  \int_{\Omega_d(H, z)} d \widehat{\bm
  \sigma}\abs{\bm{c} \cdot \widehat{ \bm a}}(\bm{c} \cdot \widehat{
  \bm a}) \\ 
 \Big[ (w \widehat{\sigma}_\parallel)^2+ (w_z
 \widehat{\sigma}_z)^2  \Big]^{1/2}({\bm 
   c}\cdot \widehat{ \bm e}_z) w_z  \widehat{\sigma}_z. 
\end{split}
\end{align}
Expressing the variable $\bm{c}$ in orthonormal basis with components
$\{    \widehat{ \bm b}_1,\dots,\widehat{ \bm b}_{d-1},\widehat{ \bm
  a}\}$ and integrating over $\bm{c}$ and $\bm{C}$, it is obtained  
\begin{align} \label{ecub16}
H_2(z) =  \frac{ 2 \sqrt{2}n^2 w^3}{\sqrt{\pi}} (1+\beta)
   \int_{\Omega_d(H, z)} \dd \widehat{\bm \sigma} \Big(1+
  \beta\widehat{\sigma}_z^2  \Big)^{1/2}   
\widehat{\sigma}_z^2.
\end{align}
Integrating over $z$ and expressing Eq. \eqref{ecub16} in polar coordinates, we have
\begin{align} \label{ecub.31}
    \begin{aligned}
    \frac{1}{(H-\sigma)} \int_{\sigma/2}^{H-\sigma/ 2} \dd z
    H_2(z)=\frac{2  \sqrt{2} n^2w^3 \Omega_{T,d-1} (1+\beta)}{
      \sqrt{\pi} 
  \varepsilon \sigma} \left[ I^{*}_1 + \sigma(\varepsilon-1)
  I^{*}_2\right] , 
\end{aligned}
\end{align}
with
\begin{align}
    I^*_1 \equiv& \int_{\sigma/2}^{3\sigma/2} \dd z
                  \int_{\pi/2-b_2(z)}^{\pi/2+ b_1(z)} \dd \theta
                  \left( 1+   \beta  \cos^2\theta  \right)^{1/2}
                  \sin^{d-2}\theta\cos^2\theta,  \label{ecu_Ip1} \\
    I^*_2 \equiv & \int_{0}^{\pi} \dd \theta \left( 1+   \beta
                   \cos^2\theta  \right)^{1/2} \sin^{d-2}\theta 
\cos^2\theta \label{ecu_Ip2}.
\end{align}
Using the same variables as above, we can express $I^*_1$ and $I^*_2$
as a function of $y$ as 
\begin{align}
    \begin{aligned}
        I^{*}_1 =& 2 \sigma \int_0^1 \mathrm{d} y
        \;y^2(1-y)\left(1-y^2\right)^{(d-3)/2} \left(1+\beta
          y^2\right)^{1 / 2}, 
    \\
     I^{*}_2 =& 2  \int_{0}^{1} \dd y  \; y^2   (1-y^2)^{(d-3)/2}
\left( 1+   \beta  y^2  \right)^{1/2}, 
     \end{aligned}
\end{align}
and, by introducing this results into Eq. \eqref{ecub.31}, we have
\begin{align} \label{ecu_H2}
    \begin{aligned}
\frac{1}{(H-\sigma)} \int_{\sigma/2}^{H-\sigma/ 2} \dd z H_2(z)=&
\frac{4  \sqrt{2} n^2w^3 \Omega_{T,d-1} (1+\beta)}{   \sqrt{\pi}\varepsilon   }
\int_0^1 \dd y \; y^2 (1-y^2)^{(d-3)/2} (1+\beta y ^2)^{1/2}
[(\varepsilon-1)+(1-y)]. 
\end{aligned}
\end{align}
Finally, by introducing Eqs. \eqref{ecu_H1} and \eqref{ecu_H2} into
Eq. \eqref{ecub3}, the particle-particle collisional contribution to 
the vertical temperature equation, i.e.  Eq. \eqref{ecu29}, is
obtained. 

Following similar steps, the corresponding contribution to the
horizontal temperature is obtained.


\begin{thebibliography}{10}

\bibitem{g03}
I. Goldhirsch, \emph{Rapid Granular Flows}, Annu. Rev. Fluid
Mech. {\bf 35}, 57 (2003). 

\bibitem{at06}
I. S. Aranson and L. S. Tsimring, \emph{Patterns and collective
  behavior in granular media: theoretical concepts},
Rev. Mod. Phys. {\bf 78}, 641 (2006). 

\bibitem{brilliantov} N. V. Brilliantov and
  T. P\"oschel,\emph{Kinetic theory of granular gases} (Oxford
  University Press on Demand, 2004). 

\bibitem{garzo19} V. Garz\'o, \emph{Granular Gaseous Flows} (Springer
  Nature, Cham, 2019). 


\bibitem{peu02}
A. Prevost, D. A. Egolf, and J. S. Urbach, \emph{Forcing and Velocity
  Correlations in a Vibrated Granular Monolayer}, Phys. Rev. Lett. {\bf 89},
084301 (2002). 

\bibitem{ou05}
J. S. Olafsen and J. S. Urbach, \emph{Two-Dimensional Melting Far from
Equilibrium in a Granular Monolayer}, Phys. Rev. Lett. {\bf 95}, 098002 
(2005).

%%  Review
\bibitem{mvprkeu05}
P. Melby, F. Vega Reyes, A. Prevost, R. Robertson, P. Kumar,
D. A. Egolf, and J. S. Urbach, \emph{The dynamics of thin vibrated
  granular layers},  J. Phys.: Condens. Matter {\bf 17}
(2005) S2689-S2704. 

\bibitem{cms12}
G. Castillo, N. M\'ujica, and R. Soto, \emph{Fluctuations and
  Criticality of a Granular Solid-Liquid-Like Phase Transition}, Phys. Rev. Lett. {\bf 109},
095701 (2012). 

\bibitem{nrtms14}
B. N\'eel, I. Rondini, A. Turzillo, N. Mujica, and R. Soto,
\emph{Dynamics of a first-order transition to an absorbing state},
Phys. Rev. E {\bf 89}, 042206 (2014). 

\bibitem{cms15}
G. Castillo, N. M\'ujica, and R. Soto, \emph{Universality and
  criticality of a second-order granular solid-liquid-like phase transition}, Phys. Rev. E {\bf 91},
012141 (2015). 

%% Review
\bibitem{gs18}
M. Guzm\'an and R. Soto, \emph{Critical phenomena in
  quasi-two-dimensional vibrated granular systems}, Phys. Rev. E {\bf 97}, 012907 (2018). 

\bibitem{bdks98}
J. J. Brey, J. W. Dufty, C. S. Kim, and A. Santos, \emph{Hydrodynamics
for a granular flow at low density}, Phys. Rev. E {\bf
  58}, 4638 (1998). 

\bibitem{brs13}
R. Brito, D. Risso, and R. Soto, \emph{Hydrodynamic modes in a
  confined granular fluid}, Phys. Rev. E {\bf 87}, 022209
(2013). 

\bibitem{mgb19}
P. Maynar, M. I. Garc\'ia de Soria, and J. J. Brey, \emph{Homogeneous
  dynamics in a vibrated granular monolayer}, J. Stat. Mech. (2019)
093205. 

\bibitem{mgb19b} P. Maynar, M. I. Garc\'ia de Soria, and J. J. Brey,
\emph{Understanding an instability in vibrated granular monolayers},
Phys. Rev. E {\bf 99}, 032903 (2019). 

\bibitem{mgb22} P. Maynar, M. I. Garc\'ia de Soria, and J. J. Brey,
\emph{Dynamics of an inelastic tagged particle under strong confinement},
Phys. Fluids {\bf 34}, 123321 (2022). 

\bibitem{bmg16} J. J. Brey, P. Maynar, M. I. Garc\'ia de Soria,
\emph{Kinetic equation and nonequilibrium entropy for a
  quasi-two-dimensional gas}, Phys. Rev. E {\bf 94}, 040103(R)
(2016). 

\bibitem{bgm17} J. J. Brey, M. I. Garc\'ia de Soria, and P. Maynar, 
\emph{Boltzmann kinetic equation for a strongly confined gas of hard
  spheres},  Phys. Rev. E {\bf 96}, 042117 (2017). 

\bibitem{mbgm22} M. Mayo, J. J. Brey, M. I. Garc\'ia de Soria, and P. Maynar, 
\emph{Kinetic Theory of a confined quasi-one-dimensional gas of hard
  disks},  Physica A {\bf 597} (2022) 127237. 

\bibitem{mgb18} P. Maynar, M. I. Garc\'ia de Soria, and J. J. Brey,
\emph{The Enskog Equation for Confined Elastic Hard Spheres},
J. Stat. Phys. {\bf 170}, 999 (2018). 

\bibitem{rcbhs11}
K. Roeller, J. P. D. Clewett, R. M. Bowley, S. Herminghaus, and
M. R. Swift, \emph{Liquid-Gas Phase Separation in Confined Vibrated
  Dry Granular Matter}, Phys. Rev. Lett. {\bf 107},
048002 (2011). 

\bibitem{crbhs12}
J. P. D. Clewett, K. Roeller, R. M. Bowley, S. Herminghaus, and
M. R. Swift, \emph{Emergent Surface Tension in Vibrated, Noncohesive
  Granular Media}, Phys. Rev. Lett. {\bf 109}, 
228002 (2012). 

\bibitem{cwbhsm16}
J. P. D. Clewett, J. Wade, R. M. Bowley, S. Herminghaus, M. R. Swift,
and M. G. Mazza, \emph{The minimization of mechanical work in vibrated
granular matter}, Sci. Rep. {\bf 6}, 28726 (2016). 


%XXXB

\bibitem{resibois1977classical} P. R\'esibois and M. de Leener, \emph{Classical kinetic theory of fluids} (Wiley, 1977).

\bibitem{dorfman2021contemporary}  J. R. Dorfman, H. van Beijeren, and T. R. Kirkpatrick, \emph{Contemporary kinetic theory of matter} (Cambridge University Press, 2021).


\bibitem{mclennan1989introduction}  J. McLennan and M. A,\emph{ Introduction to nonequilibrium statistical mechanics}
(Prentice Hall, 1989).

\bibitem{van1998velocity} T. Van Noije and M. Ernst, \emph{Velocity distributions in homogeneous granular fluids: the free and the heated case}, Granular Matter {\bf 1}, 57-64 (1998).



%\bibitem{powell1970hybrid} M. J. Powell, \emph{A hybrid method for nonlinear equations}, Numerical methods for nonlinear algebraic equations (1970).


%\bibitem{2020SciPy-NMeth} P. Virtanen, R. Gommers, T. E. Oliphant, M. Haberland, T. Reddy, D. Cournapeau, E. Burovski, P. Peterson, W. Weckesser, J. Bright, S. J. van der Walt, M. Brett, J. Wilson, K. J. Millman, N. Mayorov, A. R. J. Nelson, E. Jones, R. Kern, E. Larson, C. J. Carey, I. Polat, Y. Feng, E. W. Moore, J. VanderPlas, D. Laxalde, J. Perktold, R. Cimrman, I. Henriksen, E. A. Quintero, C. R. Harris, A. M. Archibald, A. H. Ribeiro, F. Pedregosa, P. van Mulbregt, and SciPy 1.0 Contributors, \emph{ SciPy 1.0: Fundamental Algorithms for Scientific Computing in Python}, Nature Methods {\bf 17}, 261-272 (2020).


%\bibitem{simpy} A. Meurer, C. P. Smith, M. Paprocki, O. Certík,
%S. B. Kirpichev, M. Rocklin, A. Kumar, S. Ivanov, J. K. Moore,
%S. Singh, T. Rathnayake, S. Vig, B. E. Granger, R. P. Muller,
%F. Bonazzi, H. Gupta, S. Vats, F. Johansson, F. Pedregosa,
%M. J. Curry, A. R. Terrel, Š. Roučka, A. Saboo, I. Fernando,
%S. Kulal, R. Cimrman, and A. Scopatz, \emph{Sympy: symbolic computing
%in python}, Peer. Computer Science {\bf 3}. e103(2017).

\bibitem{bgm19} J. J. Brey, M. I. Garc\'ia de Soria, and P. Maynar, 
\emph{Inhomogeneous cooling state of a strongly confined granular gas
  at low density},  Phys. Rev. E {\bf 100}, 052901 (2019). 


\bibitem{mgmTBP}
M. Mayo, M. I. Garc\'ia de Soria, and P. Maynar, to be
published. 


\bibitem{allen}
M. P. Allen and D. J. Tisdesley, \emph{Computer Simulations of
  Liquids} (Oxford Science Publications, New York, 1987). 

\bibitem{pt14}
A. Prados and E. Trizac, \emph{Kovacs-like memory effect in driven
  granular gases}, Phys. Rev. Lett. {\bf 112}, 198001 (2014). 

\bibitem{bgmb14}
J. J. Brey, M. I. Garc\'ia de Soria, P. Maynar, and V. Buz\'on, 
\emph{Memory effects in the relaxation of a confined granular gas},
Phys. Rev. E {\bf 90}, 032207 (2014).  

\bibitem{lvps17}
A. Lasanta, F. Vega Reyes, A. Prados, and A. Santos, \emph{When the
  hotter cools more quickly: Mpemba effect in granular fluids},
Phys. Rev. Lett. {\bf 119}, 148001 (2017).  

\bibitem{bpr22}
A. Biswas, V. V. Prasad, and R. Rajesh, \emph{Mpemba Effect in
  Anisotropically Driven Inelastic Maxwell Gases}, J. Stat. Phys. {\bf
  186}, 45 (2022). 

\bibitem{gmt12}
M. I. Garc\'ia de Soria, P. Maynar, and E. Trizac, \emph{Universal
  reference state in a driven homogeneous granular gas}, Phys. Rev. E
{\bf 85}, 051301 (2012). 

\bibitem{bmgb14}
J. J. Brey, P. Maynar, M. I. Garc\'ia de Soria, and V. Buz\'on, 
\emph{Homogeneous hydrodynamics of a collisional model of confined
  granular gases}, Phys. Rev. E {\bf 89}, 052209 (2014).  

%\bibitem{nagle2011fundamentals} R. K. Nagle, E. B. Saff, and A. D. Snider, Fundamentals %of differential equations and boundary value problems (Pearson education, 2011).



\end{thebibliography}
\end{document}